\documentclass[11pt]{article}

\usepackage{amsmath,amssymb,amsfonts}
\usepackage{amsthm}
\usepackage{mathrsfs}

\usepackage{graphicx}
\usepackage{pgf,tikz}
\usepackage{caption}

\usepackage{multirow}
\usepackage{booktabs}
\usepackage{algorithm}
\usepackage{algorithmicx}
\usepackage{algpseudocode}
\usepackage{listings}

\usepackage[title]{appendix}
\usepackage{xcolor}
\usepackage{textcomp}
\usepackage{manyfoot}
\usepackage[superscript]{cite}
\usepackage{lscape}
\usepackage{setspace}
\usepackage{geometry}
\usepackage{authblk}
\usepackage{hyperref}
\usepackage{lineno}

\usetikzlibrary{
    shapes,
    decorations,
    arrows,
    calc,
    arrows.meta,
    fit,
    positioning,
    shapes.arrows,
    shapes.geometric,
    shapes.multipart,
    decorations.pathmorphing,
    shapes.swigs
}

\tikzset{
    -Latex,
    auto,
    node distance = 1 cm and 1 cm,
    semithick,
    state/.style = {ellipse, draw, minimum width = 0.7 cm},
    point/.style = {circle, draw, inner sep=0.04cm, fill, node contents={}},
    bidirected/.style = {Latex-Latex, dashed},
    el/.style = {inner sep=2pt, align=left, sloped}
}

\DeclareGraphicsExtensions{.pdf, .jpg, .png}
\setkeys{Gin}{width=0.85\textwidth}

\newtheorem{theorem}{Theorem}

\newtheorem{corollary}{Corollary}

\newtheorem{definition}{Definition}

\newcommand{\E}{\mathbb{E}}
\newcommand{\ind}{\perp\!\!\!\!\perp}

\title{\LARGE On the Graphical Rules for Recovering the Average Treatment Effect Under Selection Bias}

\author[1]{Yichi Zhang}
\author[2,3]{Haidong Lu\thanks{\large Address for correspondence: Dr. Haidong Lu, Department of Internal Medicine, Yale University School of Medicine, New Haven, CT; email: \href{mailto:haidong.lu@yale.edu}{haidong.lu@yale.edu}}}

\affil[1]{Department of Biostatistics, Yale School of Public Health, New Haven, CT, USA}
\affil[2]{Department of Internal Medicine, Yale School of Medicine, New Haven, CT, USA}
\affil[3]{Department of Chronic Disease Epidemiology, Yale School of Public Health, New Haven, CT, USA}

\date{\large 7\textsuperscript{th} July, 2026}

\begin{document}
\maketitle
\thispagestyle{empty}

\newpage

\pagenumbering{arabic}

\begin{abstract}
\noindent Selection bias is a major obstacle toward valid causal inference in epidemiology. Over the past decade, several graphical rules based on causal diagrams have been proposed as sufficient identification conditions for addressing selection bias and recovering causal effects. However, these simple graphical rules are typically coupled with specific identification strategies and estimators. In this article, we show two important cases of selection bias that fall outside the scope of these existing simple rules and their estimators: one case where selection is a descendant of a collider of the treatment and the outcome, and the other case where selection is affected by a mediator. To address selection bias and recover the marginal average treatment effect in these two cases, we propose an alternative set of graphical rules and construct identification formulas using g-computation and inverse probability weighting (IPW) based on single-world intervention graphs (SWIGs), leveraging external information from individuals outside the selected sample. We conduct simulation studies to verify the performance of the estimators when the traditional crude selected-sample analysis (i.e., complete-case analysis) yields erroneous conclusions that contradict the truth. We emphasize that identifying causal effects in the presence of selection bias depends critically on both the target estimand and the availability of external information.  \\[1em]
\end{abstract}
\newpage

\section{Introduction}\label{sec1}

Selection bias is one of the fundamental obstacles to causal inference. The concept of selection bias encompasses various forms stemming from different mechanisms for individuals being selected into the study or analysis \cite{hernan2004structural}. The complexity of these selection mechanisms poses difficulty in a structural definition of selection bias, which has been continually evolving over the past two decades \cite{hernan2017invited, lu2022toward, kenah2023potential, lu2024evolution}. Recently, a refined definition of selection bias that unifies different types of selection bias was proposed \cite{lu2022toward} by considering the true causal effect in the population prior to the selection process as the reference. This definition can be decomposed into two types: type 1 selection bias due to restricting to at least one level of a collider or its descendant, and type 2 selection bias due to restricting to at least one level of an effect measure modifier.\cite{lu2022toward}

With a clearer understanding of selection bias in place, recent methodological developments have been dedicated to recovering causal effects in the presence of selection bias. Specifically, aided by causal diagrams, including directed acyclic graphs (DAGs) and single-world intervention graphs (SWIGs)\cite{richardson2013single, bezuidenhout2025swigs}, several graphical rules were proposed to assess and account for selection bias, either alone or in combination with other challenges (e.g., confounding bias or missing data), general or specific to certain study designs (e.g., case-control studies) and to particular causal estimands that vary both in the level of aggregation (conditional versus marginal average treatment effect) and in the effect measure (causal risk difference or risk ratio versus causal odds ratio)\cite{bareinboim2014recovering, correa2018generalized, mathur2025simple, schnell2024graphical, kenah2023potential}.

While these graphical rules provide valuable insights, they are typically paired with specific identification strategies or estimators. For example, under the graphical rules proposed by Mathur and Shpitser, one key premise is that post-treatment variables cannot be conditioned on when conducting selected-sample analysis (i.e., complete-case analysis) \cite{mathur2025simple}. In addition, their focus is on the conditional average treatment effect (CATE), rather than the average treatment effect (ATE) which is arguably more widely used. Other approaches, by contrast, already permit conditioning on post-treatment variables in specific settings. For example, the potential outcomes approach proposed by Kenah allows this to identify the causal odds ratio in case–control studies \cite{kenah2023potential}. Such approaches, however, do not permit conditioning on descendants of the outcome, and focus on the causal odds ratio. By contrast, epidemiologic analyses have frequently relied on inverse probability of selection or censoring weights to address different forms of selection bias (e.g., due to non-random sample selection or differential loss to follow-up). \cite{hernan2004structural, howe2016selection} These approaches leverage covariate information (potentially including post-treatment variables) from unselected or censored individuals, rather than relying solely on selected-sample analysis. Consequently, the graphical rules proposed by Mathur and Shpitser may not be universally applicable. Other existing graphical rules, while not prohibiting adjustment for post-treatment variables, impose strong sufficient conditions intended to avoid confounding bias \cite{correa2017causal, correa2018generalized, bareinboim2014recovering}. These conditions are sufficient but not necessary. They are formulated to rule out confounding under arbitrary structures, so in the settings we consider, where treatment is randomized or the relevant backdoor paths are otherwise blocked, they are not satisfied by the post-treatment adjustment that removes selection bias, even though that adjustment introduces no confounding bias. As a result, these rules may not certify a truly identifiable ATE when the primary challenge is selection bias rather than confounding. 

Given these limitations, some important cases of selection bias fall outside the scope of existing graphical rules when inference is restricted to information available within the selected sample. While it is reasonable and legitimate to focus on the selected-sample analysis, in many epidemiologic settings, additional information may be available from individuals who were not selected into the analytic sample. Such external information can expand the range of settings in which causal effects are identifiable and motivates consideration of alternative identification strategies. For example, the available external information may consist of post-treatment variables observed among unselected individuals \cite{breskin2018practical}. Notably, this may require only the distribution of some post-treatment variables in the full population. The outcome itself may remain unobserved among the unselected, so the relaxation is limited and its requirements are made explicit. In such cases, identification of causal effects (particularly the ATE) may still be possible using estimators such as g-computation \cite{breskin2018practical, zivich2025gcomp}, even though identification would be ruled out by existing graphical rules. This underscores the value of applying fundamental graphical principles to guide the search for appropriate identification strategies. For instance, inverse probability weighting (IPW) and g-computation remain powerful tools for recovering treatment effects when external information beyond the selected sample is available \cite{hernan2004structural, correa2018generalized, breskin2018practical}.

In this article, we examine two important cases of selection bias that are not addressed by existing graphical rules. The first arises when conditioning on a descendant of a collider between the exposure and the outcome. The second occurs when sample selection is affected by a mediator. We show that in both cases, the ATE can be recovered using g-computation or IPW, provided that external covariate information from the unselected sample is available. We formally state the conditions under which these identification strategies hold, and demonstrate how they can be applied to address selection bias in related settings. In addition, we propose two IPW estimators for the ATE, along with corresponding variance estimators, and evaluate their statistical properties through simulation studies. In our hypothetical simulation settings, we find that conventional complete-case analyses based solely on the selected sample yield conclusions that contradict the true causal effects and risk misleading decision-making. Finally, we emphasize the importance of clearly specifying the target estimand and the availability of external information when identifying causal effects in the presence of selection bias.

We note that the settings we consider correspond to a special case of monotone missing-at-random (MAR), in which the outcome is the only incomplete variable and the fully observed post-treatment variable d-separates the outcome from selection \cite{seaman2013meant, sun2018inverse, mohan2021graphical}. Our contribution is therefore not a new identification result for MAR, for which an extensive literature exists, but a graphical one: from the perspective of the existing graphical rules for selection bias, these cases appear unidentifiable, yet applying the fundamental d-separation principles on SWIGs recovers the marginal ATE, and in these cases the selection-bias and missing-data perspectives coincide. Notably, this licenses conditioning on a post-treatment variable, even a descendant of the outcome (a collider) or a mediator, which the prevailing graphical rules for selection bias would exclude.

\section{Two cases of selection bias unaddressed by simple graphical rules}\label{sec2}

To detect and control selection bias, researchers often rely on simple graphical rules in conjunction with a specific class of identification formulas, and define identifiability (or recoverability) narrowly as whether the treatment effect of interest can be identified within that class \cite{correa2017causal,correa2018generalized,bareinboim2014recovering,mathur2025simple}. These graphical rules typically provide sufficient conditions for establishing that the treatment effect is identifiable (or recoverable) using a given set of formulas. However, failure to satisfy these conditions does not imply that the effect is unidentifiable; rather, it indicates that identification may require alternative strategies or estimators. In this section, we present two important cases of selection bias that fall outside the scope of previously proposed graphical rules. Identification of treatment effects for the related causal structures using modified formulas is discussed in later sections.

\subsection{Setting}
\begin{figure}
\centering
\begin{tikzpicture}

% DAG for Case 1
% nodes %
\node (A_DAG1) {$A$};
\node[right = 3 of A_DAG1, text centered] (Y_DAG1) {$Y$};
\node[below right = 1.7 of A_DAG1, text centered] (L_DAG1) {$L$};
\node[below = 1 of L_DAG1, text centered] (S_DAG1) {$S$};
% edges %
\draw[->] (A_DAG1) --  (Y_DAG1);
\draw[->] (A_DAG1) -- (L_DAG1);
%\draw[->] (A_DAG1) -- (S_DAG1);
\draw[->] (Y_DAG1) -- (L_DAG1);
\draw[->] (L_DAG1) -- (S_DAG1);

% SWIG for Case 1
% nodes %
\node[name=A_SWIG1, right = 1 of Y_DAG1, text centered, shape=swig vsplit]{
\nodepart{left}{$A$}
\nodepart{right}{$a$} };
\node[right = 3 of A_SWIG1, text centered] (Y_SWIG1) {$Y(a)$};
\node[below right = 1.7 of A_SWIG1, text centered] (L_SWIG1) {$L(a)$};
\node[below = 1 of L_SWIG1, text centered] (S_SWIG1) {$S(a)$};
% edges %
\draw[->] (A_SWIG1) --  (Y_SWIG1);
\draw[->] (A_SWIG1) -- (L_SWIG1);
%\draw[->] (A_SWIG1) -- (S_SWIG1);
\draw[->] (Y_SWIG1) -- (L_SWIG1);
\draw[->] (L_SWIG1) -- (S_SWIG1);

% DAG for Case 2
% nodes %
\node[below = 5 of A_DAG1, text centered] (A_DAG2) {$A$};
\node[right = 1.2 of A_DAG2, text centered] (L_DAG2) {$L$};
\node[right = 1.2 of L_DAG2, text centered] (Y_DAG2) {$Y$};
\node[below = 1.5 of L_DAG2, text centered] (S_DAG2) {$S$};
% edges %
\draw[->] (A_DAG2) to [out=60,in=120] (Y_DAG2);
\draw[->] (A_DAG2) -- (L_DAG2);
%\draw[->] (A_DAG2) -- (S_DAG2);
\draw[->] (L_DAG2) -- (Y_DAG2);
\draw[->] (L_DAG2) -- (S_DAG2);

% SWIG for Case 2
% nodes %
% nodes %
\node[name=A_SWIG2, right = 1 of Y_DAG2, text centered, shape=swig vsplit]{
\nodepart{left}{$A$}
\nodepart{right}{$a$}};
\node[right = 1.2 of A_SWIG2, text centered] (L_SWIG2) {$L(a)$};
\node[right = 1.2 of L_SWIG2, text centered] (Y_SWIG2) {$Y(a)$};
\node[below = 1.5 of L_SWIG2, text centered] (S_SWIG2) {$S(a)$};
% edges %
\draw[->] (A_SWIG2) to [out=60,in=120] (Y_SWIG2);
\draw[->] (A_SWIG2) -- (L_SWIG2);
%\draw[->] (A_SWIG2) -- (S_SWIG2);
\draw[->] (L_SWIG2) -- (Y_SWIG2);
\draw[->] (L_SWIG2) -- (S_SWIG2);

% labels %
\node[left = 1 of A_DAG1, text centered] (c1) {\textbf{Case 1}};
\node[left = 1 of A_DAG2, text centered] (c2) {\textbf{Case 2}};
\node[above = 2.2 of L_DAG1, text centered] (DAG) {\textbf{DAG}};
\node[above = 2.2 of L_SWIG1, text centered] (SWIG) {\textbf{SWIG}};

\end{tikzpicture}
\vspace*{5mm}
\caption{Causal directed acyclic graphs (DAGs) and single-world intervention graphs (SWIGs) for Case 1 and Case 2 of sample selection directly affected by a post-treatment variable $L$. When a unit is selected into the sample (i.e. $S_i = 1$), we observe all variables $(A_i, Y_i, L_i)$. When a unit is not selected (i.e. $S_i = 0$), we may observe a part or none of the variables. In Case 1, the direct upstream cause of selection, $L$, is the common effect (i.e., collider) of both the treatment $A$ and the outcome $Y$. In Case 2, the direct upstream cause of selection, $L$, is the mediator of the treatment $A$ on the outcome $Y$.}
\label{fig:DAG_SWIG}
\end{figure}
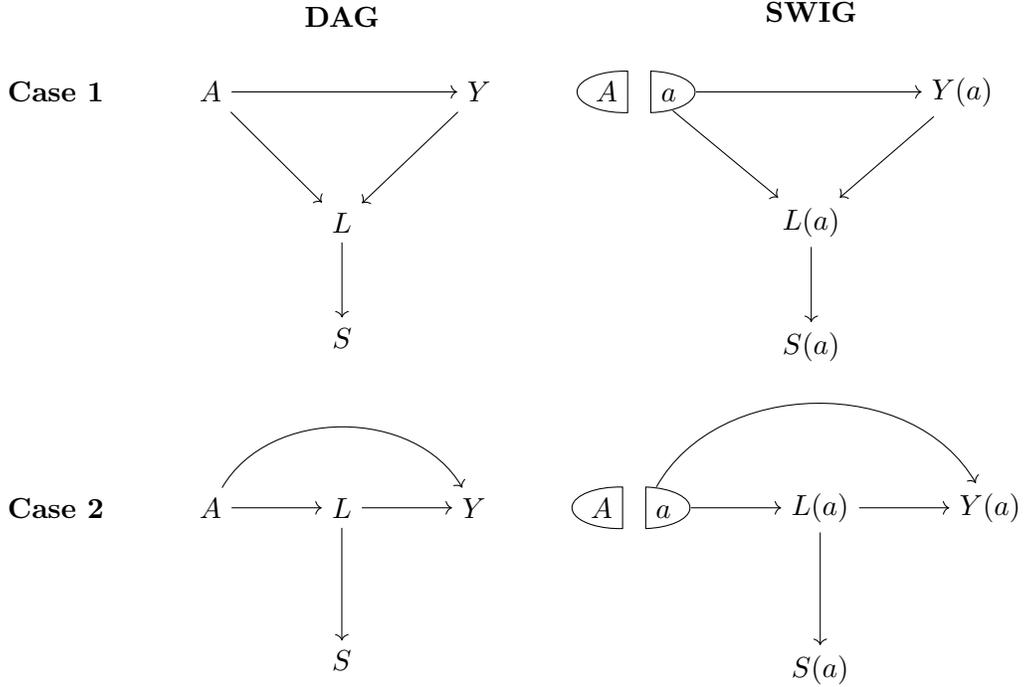

Consider $n$ individuals (hereafter called ``units") in a randomized clinical trial whose outcomes are observed only among selected units. The selection is directly affected by a post-treatment variable observable among all units. Specifically, let $A_i$ be the binary treatment for unit $i$, $Y_i$ be the binary outcome, and $S_i$ be the selection. There are no other variables as confounders of $A_i$ and $Y_i$. We define the direct upstream cause of the selection to be $L_i$, which can be either a common effect of the treatment and the outcome (Case 1 in Fig. \ref{fig:DAG_SWIG}), or a mediator of the treatment on the outcome (Case 2 in Fig. \ref{fig:DAG_SWIG}). When a unit is selected, i.e. $S_i = 1$, we observe all variables $(A_i, Y_i, L_i)$. When a unit is not selected, i.e. $S_i = 0$, we possibly observe a part or none of the variables. Under the consistency assumption, $(Y_i(a_i), L_i(a_i), S_i(a_i))$ = $(Y_i, L_i, S_i)$.

Assume that our goal is to recover the average treatment effect (ATE) in the general population prior to the selection process:
\begin{equation*}
    \delta_{\text{ATE}} = \E[Y(1)-Y(0)].
\end{equation*}
We focus on causal contrasts, and specifically on the causal risk difference. Other effect measures such as the causal risk ratio or odds ratio can also be derived, and type 2 selection bias may occur on one scale but not the other, so the risk ratio can be unbiased where the risk difference is not, and vice versa \cite{lu2022toward}. The odds ratio is further complicated by non-collapsibility.

%%%%%%%%%%%%%%%%%%DAGs and SWIGs%%%%%%%%%%%%%%%%%%%%%%
%%%%%%%%%%%%%%%%%%DAGs and SWIGs%%%%%%%%%%%%%%%%%%%%%%

\subsection{Limitations of existing simple graphical rules}

We focus on three existing graphical rules proposed over the past decade: the selection-backdoor criterion \cite{bareinboim2014recovering}, the generalized adjustment criterion \cite{correa2018generalized}, and the graphical rules of Mathur and Shpitser \cite{mathur2025simple}. In Appendix \ref{secA1}, we demonstrate that the two cases discussed above violate these rules, and that identification by their corresponding formulas is therefore not guaranteed.

\section{Identification by the fundamentals of graphical rules with external data}\label{sec3}

%Existing simple graphical rules are derived as compared to and based on the fundamentals of DAGs and SWIGs \cite{pearl2000models, richardson2013single}. The core element is to read off conditional independence based on the fundamental definition of d-separation that invites an open-ended search for an identification formula. When the graphical structure of the problem falls outside of the expertise by the existing simple graphical rules, it naturally invites an ad-hoc investigation of the identifiability and possible construction of unconventional formulas. If the ATE can be proved to be identified by some other formulas the conditions that validate the identification can be extracted as new graphical rules for these formulas. Their utility depends on how often the stipulated conditions hold under realistic circumstances.

In this section, we first introduce a g-computation formula that can recover the ATE by conditioning on post-treatment variables \cite{breskin2018practical} using external data (i.e., covariate information from the unselected sample). We then propose IPW estimators to recover the ATE and address selection bias in the two cases described above. The identification results from these methods are further generalized under specific graphical rules to allow (i) observed confounders between the treatment and post-treatment variables, (ii) unobserved confounders between post-treatment variables and the outcome, and (iii) selection processes directly affected by the treatment.

\subsection{G-computation formulas}

In Breskin et al. \cite{breskin2018practical}, a special-case g-formula conditioning on a post-treatment variable $L$ was derived to identify the ATE in a randomized controlled trial example, illustrating the utility of SWIGs. We extend their results to observational settings by allowing for the presence of confounders. Our identification results hold when external information on covariates among unselected units is available, together with knowledge of $A_i$ or the treatment assignment mechanism $p(A_i = 1)$, while permitting the outcome $Y_i$ to remain unobserved.

Specifically, we propose the following theorem.
\begin{theorem}[Identification of the ATE by the g-formula under selection bias]
\label{thrm:g_formula}
Given a SWIG $\mathcal{G}(a)$ containing the random treatment component $A$, the fixed treatment component $a$, the post-treatment variables $L(a)$, the selection variable $S(a)$, the outcome variable $Y(a)$, the observed pre-treatment variables $X$, and the unobserved pre-treatment variables $U$, the ATE is identified by
\begin{equation}
\label{g_formula}
\begin{split}
        \delta_{\text{ATE}, g} =  \sum_x (\sum_l &\E[Y\,|\,A=1, L=l, S=1, X=x]p(L=l\,|\,A=1, X=x) -\\
        &\E[Y\,|\,A=0, L=l, S=1, X=x]p(L=l\,|\,A=0, X=x))p(X=x),
\end{split}
\end{equation}
i.e. $\delta_{\text{ATE}} = \delta_{\text{ATE}, g}$, if the following conditions hold:
\begin{equation}
\label{g_formula_condition}
\begin{split}
    &(C1) \quad Y(a) \ind S(a) \,|\, \{L(a), X\}\\
    &(C2) \quad Y(a) \ind A \,|\, \{L(a), S(a), X\}\\
    &(C3) \quad L(a) \ind A \,|\, X.
\end{split}
\end{equation}
\end{theorem}
Intuitively, the first condition assumes that all paths between $Y(a)$ and $S(a)$ are blocked by $L(a)$ and $X$. This motivates the adjustment of some post-treatment variables $L(a)$ if it helps separate the outcome from the selection. The next two conditions assume that the effect of the treatment on both $Y(a)$ and $L(a)$ is unconfounded by conditioning on different sets of variables.

Therefore, despite being motivated by an ad-hoc estimator for a special case, the g-formula (\ref{g_formula}) will recover the ATE when some post-treatment
variables can be conditioned on to address selection bias. The terms $p(L=l\,|\,A=a, X=x)$ and $p(X=x)$ are not conditioned on $S = 1$ and therefore require external information from the unselected sample. The conditions can be viewed as derived graphical rules in this specific context. The example by Breskin et al. \cite{breskin2018practical} is a special case of our theorem by the following corollary.
\begin{corollary}  
[Identification of the ATE by the g-formula given SWIG \cite{breskin2018practical}]
\label{crlr:g_formula_breskin}
The ATE is identified by 
\begin{equation*}
\begin{split}
        \delta_{\text{ATE}, g} =  \sum_l &\E[Y\,|\,A = 1, L = l, S = 1]p(L=l\,|\,A=1) -\\
        &\E[Y\,|\,A = 0, L = l, S = 1]p(L=l\,|\,A=0),
\end{split}
\end{equation*}
i.e. $\delta_{\text{ATE}} = \delta_{\text{ATE}, g}$, for the SWIG in Breskin et al.\cite{breskin2018practical}.
\end{corollary}

Our Case 2 can be considered as a modification of the graph in Breskin et al. \cite{breskin2018practical} where the derivation of the formula still applies. Moreover, it is straightforward to verify that the conditions (\ref{g_formula_condition}) are satisfied in both of our cases by the SWIGs in Fig. \ref{fig:DAG_SWIG}. We have the following corollary as examples.
\begin{corollary}  
[Identification of the ATE by the g-formula given Fig. \ref{fig:DAG_SWIG}]
\label{crlr:g_formula_Fig1}
The ATE is identified by
\begin{equation*}
\begin{split}
        \delta_{\text{ATE}, g} =  \sum_l &\E[Y\,|\,A = 1, L = l, S = 1]p(L=l\,|\,A=1) -\\
        &\E[Y\,|\,A = 0, L = l, S = 1]p(L=l\,|\,A=0),
\end{split}
\end{equation*}
i.e. $\delta_{\text{ATE}} = \delta_{\text{ATE}, g}$, for the two SWIGs in Fig. \ref{fig:DAG_SWIG}.
\end{corollary}

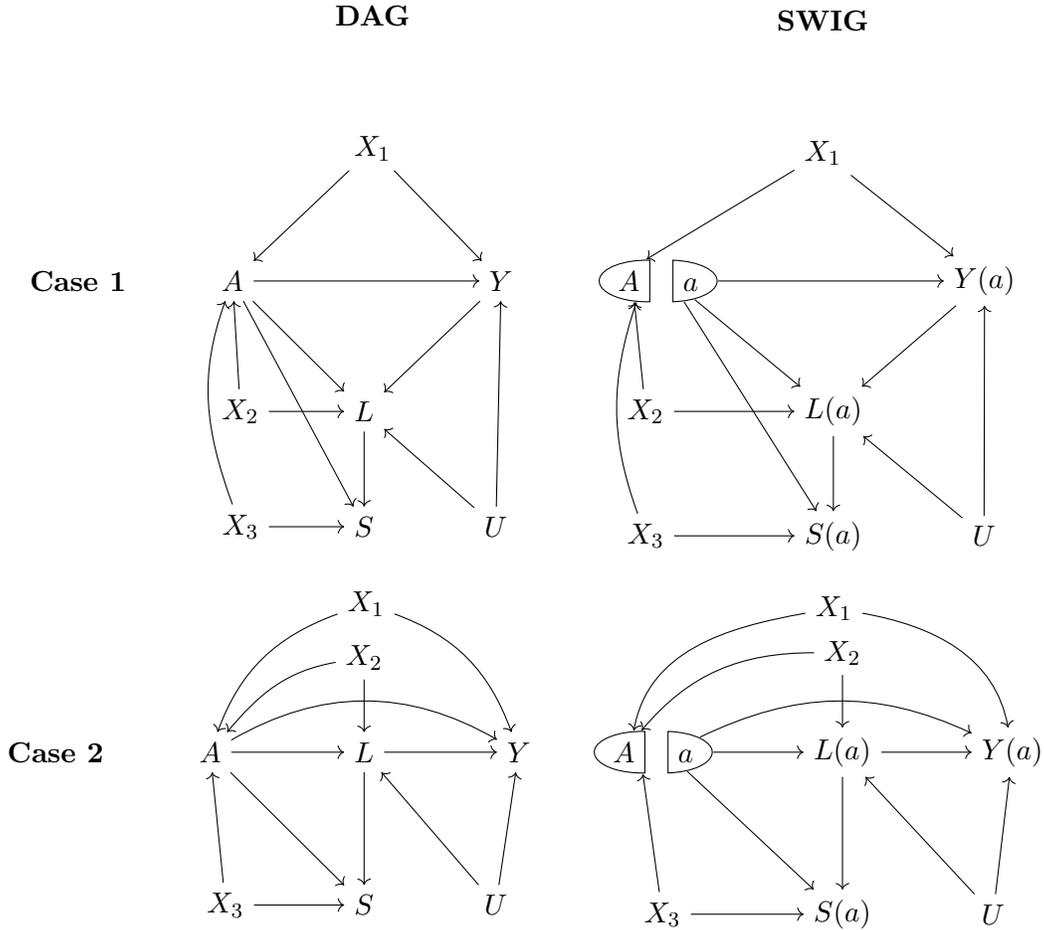
\begin{figure}
\centering
\begin{tikzpicture}

% DAG for Case 1
% nodes %
\node (A_DAG1) {$A$};
\node[right = 3 of A_DAG1, text centered] (Y_DAG1) {$Y$};
\node[below right = 1.7 of A_DAG1, text centered] (L_DAG1) {$L$};
\node[below = 1 of L_DAG1, text centered] (S_DAG1) {$S$};
\node[above right = 1.7 of A_DAG1, text centered] (X1_DAG1) {$X_1$};
\node[left = 1 of L_DAG1, text centered] (X2_DAG1) {$X_2$};
\node[left = 1 of S_DAG1, text centered] (X3_DAG1) {$X_3$};
\node[right = 1.2 of S_DAG1, text centered] (U_DAG1) {$U$};
% edges %
\draw[->] (A_DAG1) -- (Y_DAG1);
\draw[->] (A_DAG1) -- (L_DAG1);
\draw[->] (A_DAG1) -- (S_DAG1);
\draw[->] (Y_DAG1) -- (L_DAG1);
\draw[->] (L_DAG1) -- (S_DAG1);
\draw[->] (X1_DAG1) -- (A_DAG1);
\draw[->] (X1_DAG1) -- (Y_DAG1);
\draw[->] (X2_DAG1) -- (A_DAG1);
\draw[->] (X2_DAG1) -- (L_DAG1);
\draw[->] (X3_DAG1) to [out=110,in=250] (A_DAG1);
\draw[->] (X3_DAG1) -- (S_DAG1);
\draw[->] (U_DAG1) -- (L_DAG1);
\draw[->] (U_DAG1) -- (Y_DAG1);

% SWIG for Case 1
% nodes %
\node[name=A_SWIG1, right = 1 of Y_DAG1, text centered, shape=swig vsplit]{
\nodepart{left}{$A$}
\nodepart{right}{$a$} };
\node[right = 3 of A_SWIG1, text centered] (Y_SWIG1) {$Y(a)$};
\node[below right = 1.7 of A_SWIG1, text centered] (L_SWIG1) {$L(a)$};
\node[below = 1 of L_SWIG1, text centered] (S_SWIG1) {$S(a)$};
\node[above right = 1.7 of A_SWIG1, text centered] (X1_SWIG1) {$X_1$};
\node[left = 1.6 of L_SWIG1, text centered] (X2_SWIG1) {$X_2$};
\node[left = 1.6 of S_SWIG1, text centered] (X3_SWIG1) {$X_3$};
\node[right = 1.2 of S_SWIG1, text centered] (U_SWIG1) {$U$};
% edges %
\draw[->] (A_SWIG1) -- (Y_SWIG1);
\draw[->] (A_SWIG1) -- (L_SWIG1);
\draw[->] (A_SWIG1) -- (S_SWIG1);
\draw[->] (Y_SWIG1) -- (L_SWIG1);
\draw[->] (L_SWIG1) -- (S_SWIG1);
\draw[->] (X1_SWIG1) -- (A_SWIG1.450);
\draw[->] (X1_SWIG1) -- (Y_SWIG1);
\draw[->] (X2_SWIG1) -- (A_SWIG1.600);
\draw[->] (X2_SWIG1) -- (L_SWIG1);
\draw[->] (X3_SWIG1) to [out=110,in=250] (A_SWIG1);
\draw[->] (X3_SWIG1) -- (S_SWIG1);
\draw[->] (U_SWIG1) -- (L_SWIG1);
\draw[->] (U_SWIG1) -- (Y_SWIG1);

% DAG for Case 2
% nodes %
\node[below = 4 of L_DAG1, text centered] (L_DAG2) {$L$};
\node[left = 1.5 of L_DAG2, text centered] (A_DAG2) {$A$};
\node[right = 1.5 of L_DAG2, text centered] (Y_DAG2) {$Y$};
\node[below = 1.5 of L_DAG2, text centered] (S_DAG2) {$S$};
\node[above right = 2 of A_DAG2, text centered] (X1_DAG2) {$X_1$};
\node[above = 0.7 of L_DAG2, text centered] (X2_DAG2) {$X_2$};
\node[left = 1.2 of S_DAG2, text centered] (X3_DAG2) {$X_3$};
\node[right = 1.2 of S_DAG2, text centered] (U_DAG2) {$U$};
% edges %
\draw[->] (A_DAG2) to [out=30,in=150] (Y_DAG2);
\draw[->] (A_DAG2) -- (L_DAG2);
\draw[->] (A_DAG2) -- (S_DAG2);
\draw[->] (L_DAG2) -- (Y_DAG2);
\draw[->] (L_DAG2) -- (S_DAG2);
\draw[->] (X1_DAG2) to [out=200,in=70]  (A_DAG2);
\draw[->] (X1_DAG2) to [out=340,in=110]  (Y_DAG2);
\draw[->] (X2_DAG2) to [out=190,in=50]  (A_DAG2);
\draw[->] (X2_DAG2) --  (L_DAG2);
\draw[->] (X3_DAG2) --  (A_DAG2);
\draw[->] (X3_DAG2) --  (S_DAG2);
\draw[->] (U_DAG2) --  (L_DAG2);
\draw[->] (U_DAG2) --  (Y_DAG2);

% SWIG for Case 2
% nodes %
% nodes %
\node[name=A_SWIG2, right = 0.7 of Y_DAG2, text centered, shape=swig vsplit]{
\nodepart{left}{$A$}
\nodepart{right}{$a$}};
\node[right = 1.2 of A_SWIG2, text centered] (L_SWIG2) {$L(a)$};
\node[right = 1.2 of L_SWIG2, text centered] (Y_SWIG2) {$Y(a)$};
\node[below = 1.5 of L_SWIG2, text centered] (S_SWIG2) {$S(a)$};
\node[above right = 2 of A_SWIG2, text centered] (X1_SWIG2) {$X_1$};
\node[above = 0.7 of L_SWIG2, text centered] (X2_SWIG2) {$X_2$};
\node[left = 1.5 of S_SWIG2, text centered] (X3_SWIG2) {$X_3$};
\node[right = 1.2 of S_SWIG2, text centered] (U_SWIG2) {$U$};
% edges %
\draw[->] (A_SWIG2) to [out=25,in=155] (Y_SWIG2);
\draw[->] (A_SWIG2) -- (L_SWIG2);
\draw[->] (A_SWIG2) -- (S_SWIG2);
\draw[->] (L_SWIG2) -- (Y_SWIG2);
\draw[->] (L_SWIG2) -- (S_SWIG2);
\draw[->] (X1_SWIG2) to [out=190,in=80] (A_SWIG2.470);
\draw[->] (X1_SWIG2) to [out=350,in=100]  (Y_SWIG2);
\draw[->] (X2_SWIG2) to [out=180,in=50] (A_SWIG2.450);
\draw[->] (X2_SWIG2) -- (L_SWIG2);
\draw[->] (X3_SWIG2) --  (A_SWIG2);
\draw[->] (X3_SWIG2) --  (S_SWIG2);
\draw[->] (U_SWIG2) --  (L_SWIG2);
\draw[->] (U_SWIG2) --  (Y_SWIG2);

% labels %
\node[left = 1 of A_DAG1, text centered] (c1) {\textbf{Case 1}};
\node[left = 1 of A_DAG2, text centered] (c2) {\textbf{Case 2}};
\node[above = 1.2 of X1_DAG1, text centered] (DAG) {\textbf{DAG}};
\node[above = 1.2 of X1_SWIG1, text centered] (SWIG) {\textbf{SWIG}};

\end{tikzpicture}
\vspace*{5mm}
\caption{Causal directed acyclic graphs (DAGs) and single-world intervention graphs (SWIGs) for Case 1 and Case 2 of sample selection directly affected by a post-treatment variable $L$ with observed covariates $X = (X_1, X_2, X_3)$ and unobserved covariates $U$. When a unit is selected (i.e. $S_i = 1$), we observe all variables $(X_i, A_i, Y_i, L_i)$. When a unit is not selected (i.e. $S_i = 0$), we may observe a part or none of the variables. Case 1: The direct upstream cause of selection, $L$, is the common effect (i.e., collider) of the treatment $A$ and the outcome $Y$. Case 2: The direct upstream cause of selection, $L$, is the mediator on the causal pathway from treatment $A$ to the outcome $Y$.}
\label{fig:DAG_SWIG_extension}
\end{figure}

The theorem implies that the ATE can still be recovered if we extend the two cases in Fig. \ref{fig:DAG_SWIG} to allow observed confounders $X = (X_1, X_2, X_3)$ for $A$ and $(Y, L, S)$ (i.e., treatment-post-treatment-variable confounders) in the full population, respectively, and unobserved confounders $U$ for $L$ and $Y$ (i.e., collider-outcome confounders, or mediator-outcome confounders) as in Fig. \ref{fig:DAG_SWIG_extension}.
\begin{corollary}  
[Identification of the ATE by the g-formula given Fig. \ref{fig:DAG_SWIG_extension}]
\label{crlr:g_formula_Fig2}
The ATE is identified by
\begin{equation*}
\begin{split}
        \delta_{\text{ATE}, g} =  \sum_x (\sum_l &\E[Y\,|\,A=1, L=l, S=1, X=x]p(L=l\,|\,A=1, X=x) -\\
        &\E[Y\,|\,A=0, L=l, S=1, X=x]p(L=l\,|\,A=0, X=x))p(X=x),
\end{split}
\end{equation*}
i.e. $\delta_{\text{ATE}} = \delta_{\text{ATE}, g}$, for the two SWIGs in Fig. \ref{fig:DAG_SWIG_extension}.
\end{corollary}

Note that the ATE is identified even when $X = (X_1, X_2)$. In other words, the treatment-selection confounders, $X_3$, are not required to be observed for recovering the ATE for the two cases in Fig. \ref{fig:DAG_SWIG_extension} by this g-formula. The intuition is that, if there exists some post-treatment variable $L(a)$ as the direct cause of selection that helps separate $Y(a)$ from $S(a)$, the selection $S(a)$ becomes a collider on the backdoor paths from the treatment $A$ to $L(a)$ and $Y(a)$ that go through $X_3$. Therefore, these backdoor paths ($A \leftarrow X_3 \rightarrow S(a) \leftarrow L(a)$ and $A \leftarrow X_3 \rightarrow S(a) \leftarrow L(a) \leftarrow Y(a)$) are blocked without the need to adjust for $X_3$ to correct confounding bias.

Proofs of Theorem \ref{thrm:g_formula} and its corollaries are given in Appendix \ref{secA2}.

For estimation and inference by the g-formulas, plug-in estimators can be applied and the variances can be derived using the delta method or bootstrap \cite{imai2010general}.

%%%%%%%%%%%%%%%%%%DAGs and SWIGs%%%%%%%%%%%%%%%%%%%%%%

\subsection{Inverse probability weighting}

Inverse probability weighting (IPW) methods intuitively create a pseudo-population where the distributions of covariates are balanced across treatment groups to fix the endogeneity of treatment and recover its effects \cite{rosenbaum1983central, austin2015moving}. They have also been recognized and applied for the generalizability and transportability of study results to a target population of interest that is different from the study sample  \cite{cole2010generalizing, stuart2011use, lesko2017generalizing, dahabreh2019generalizing, dahabreh2020extending, degtiar2023review}.

In the context of selection bias, if external information on the distribution of the post-treatment variable $L$, is available, it is also possible to reweight the distributions of post-treatment variables such that they are not skewed by sample selection among the pseudo-population, thereby addressing selection bias. We propose a theorem to formalize the idea.
\begin{theorem}[Identification of the ATE by the IPW method under selection bias]
\label{thrm:IPW_HT}
Given a SWIG $\mathcal{G}(a)$ containing the random treatment component $A$, the fixed treatment component $a$, the post-treatment variables $L(a)$, the selection variable $S(a)$, the outcome variable $Y(a)$, the observed pre-treatment variables $X$, and the unobserved pre-treatment variables $U$, the ATE is identified by
\begin{equation}
\label{IPW_HT}
\begin{split}
        \delta_{\text{ATE}, \text{IPW}} =  \E\left[\frac{ASY}{p(S = 1\,|\,L, X, A)p(A=1\,|\,X)}\right] - \E\left[\frac{(1-A)SY}{p(S = 1\,|\,L, X, A)p(A=0\,|\,X)}\right],
\end{split}
\end{equation}
i.e. $\delta_{\text{ATE}} = \delta_{\text{ATE}, \text{IPW}}$, if the following conditions hold:
\begin{equation}
\label{IPW_HT_condition}
\begin{split}
    &(C1) \quad Y(a) \ind S(a) \,|\, \{L(a), X\}\\
    &(C2') \quad (Y(a), S(a)) \ind A \,|\, \{L(a), X\}\\
    &(C3) \quad L(a) \ind A \,|\, X\\
\end{split}
\end{equation}
\end{theorem} 
The first condition assumes that all paths between $Y(a)$ and $S(a)$ are blocked by $L(a)$ and $X$. The next two conditions assume that the treatment and the potential responses of post-treatment variables (including the outcome, covariates, and selection) can be sequentially separated. The first and third conditions (C1 and C3) for the IPW method (\ref{IPW_HT_condition}) are identical to those for the g-formula (\ref{g_formula_condition}), whereas the second condition differs: C2$'$ requires the joint independence $(Y(a), S(a)) \ind A \,|\, \{L(a), X\}$, which is stronger than C2 in Theorem \ref{thrm:g_formula}, $Y(a) \ind A \,|\, \{L(a), S(a), X\}$. The requirement by the IPW method that $S(a)$ should be conditionally separated from $A$ is due to the inclusion of the inverse probability weighting model for the selection to recover the ATE.

The ATE can be recovered through the IPW method in all causal graphs in Breskin et al.\cite{breskin2018practical}, Fig. \ref{fig:DAG_SWIG}, and Fig. \ref{fig:DAG_SWIG_extension}. They are special cases of Theorem \ref{thrm:IPW_HT} by the following corollaries.
\begin{corollary}  
[Identification of the ATE by the IPW method given SWIG \cite{breskin2018practical}]
\label{crlr:IPW_HT_breskin}
The ATE is identified by 
\begin{equation*}
\begin{split}
        \delta_{\text{ATE}, \text{IPW}} =  \E\left[\frac{ASY}{p(S = 1\,|\,W)p(A=1)}\right] - \E\left[\frac{(1-A)SY}{p(S = 1\,|\,W)p(A=0)}\right],
\end{split}
\end{equation*}
i.e. $\delta_{\text{ATE}} = \delta_{\text{ATE}, \text{IPW}}$, for the SWIG in Breskin et al.\cite{breskin2018practical}.
\end{corollary}

\begin{corollary}  
[Identification of the ATE by the IPW method given Fig. \ref{fig:DAG_SWIG}]
\label{crlr:IPW_HT_Fig1}
The ATE is identified by
\begin{equation*}
\begin{split}
        \delta_{\text{ATE}, \text{IPW}} =  \E\left[\frac{ASY}{p(S = 1\,|\,L)p(A=1)}\right] - \E\left[\frac{(1-A)SY}{p(S = 1\,|\,L)p(A=0)}\right],
\end{split}
\end{equation*}
i.e. $\delta_{\text{ATE}} = \delta_{\text{ATE}, \text{IPW}}$, for the two SWIGs in Fig. \ref{fig:DAG_SWIG}.
\end{corollary}

\begin{corollary}  
[Identification of the ATE by the IPW method given Fig. \ref{fig:DAG_SWIG_extension}]
\label{crlr:IPW_HT_Fig2}
The ATE is identified by
\begin{equation*}
\begin{split}
        \delta_{\text{ATE}, \text{IPW}} =  \E\left[\frac{ASY}{p(S = 1\,|\,L, X, A)p(A=1\,|\,X)}\right] - \E\left[\frac{(1-A)SY}{p(S = 1\,|\,L, X, A)p(A=0\,|\,X)}\right],
\end{split}
\end{equation*}
i.e. $\delta_{\text{ATE}} = \delta_{\text{ATE}, \text{IPW}}$, for the two SWIGs in Fig. \ref{fig:DAG_SWIG_extension}.
\end{corollary}

The treatment-selection confounders $X_3$ cannot be omitted by the IPW method for recovering the ATE because the selection model requires adjustment for all variables that influence sample selection, including $X_3$, to ensure correct causal model specification. This requirement differs from that of the g-formula for the same graph.

Proofs of Theorem \ref{thrm:IPW_HT} and its corollaries are given in Appendix \ref{secA2}.

\section{Estimation and inference by IPW estimators}\label{section 4}

Based on the identification result in Theorem \ref{thrm:IPW_HT}, an unbiased estimator of the ATE is the Horvitz-Thompson (HT) estimator. We also propose an alternative IPW estimator, the Hájek estimator, which is a biased estimator of the ATE but reduces variance compared to the HT estimator \cite{hirano2003efficient}. The HT estimator normalizes by the total sample size, while the Hájek estimator normalizes by the sum of the weights (i.e., the sum of the inverse probabilities), rescaling the weights to sum to one.
%It is equivalent to the estimated slope from fitting a weighted least square regression of the outcome $Y$ using the treatment $A$ among all selected samples by the estimated probability of selection, $\hat{p}(S_i = 1|L_i)$. Despite that, the variance estimate of the regression slope needs to account for the uncertainty in estimating the weights in the first stage.
We propose closed-form sandwich estimators of their variances to enable statistical inference of the ATE \cite{lunceford2004stratification}. We use the two cases in Fig. \ref{fig:DAG_SWIG} as demonstrative examples by Corollary \ref{crlr:IPW_HT_Fig1}. We outline the steps of estimation and inference in Appendix \ref{secA3}.

\section{Simulation}\label{sec5}

We conduct simulation studies to verify the performance of our IPW estimators in both cases of Fig. \ref{fig:DAG_SWIG}. The null hypothesis of the treatment effect is set to be true in the first case, but not in the second case. For each case, we generate datasets with varying sample sizes of $n = 100, n = 1000, n = 10000$. The two IPW estimators are compared to the crude selected-sample analysis which only includes the observed treatment $A$ in a linear probability model of the observed outcome $Y$. In each setting, we simulate $K = 10000$ datasets and fit the models. We show that the crude analysis yields biased and misleading conclusions but our IPW methods accurately recover the true effect.

In Case 1, consider a randomized trial of a prophylactic medication for malaria among susceptible individuals. The treatment ($A$) is a binary variable indicating whether an individual is randomly assigned to the medication or not. The outcome of interest ($Y$) indicates whether malaria occurs, which is detected by a blood test. The ground truth is that the medication has a null effect on the outcome. However, both the medication and malaria increase the risk of fever. Whether fever occurs ($L$) is reported by these individuals. It affects participants' willingness to receive the blood test for malaria ($S$). Those who have a fever are more likely to receive the blood test, while those who do not have a fever feel less motivated to be tested. Specifically, we simulate observed data from the following distributions:
\begin{equation*}
\begin{split}
    &A_i \sim Bernoulli(0.5)\\
    &Y_i\,|\,A_i = a_i \sim Bernoulli(0.4)\\
    &L_i\,|\,A_i = a_i, Y_i = y_i \sim Bernoulli(0.1 + 0.3a_i + 0.5y_i)\\
    &S_i\,|\,L_i = l_i \sim Bernoulli(0.1 + 0.8l_i).
\end{split}
\end{equation*}
The true ATE of the treatment is $0$, meaning that the medication on average has a zero impact on reducing the risk of malaria.

\begin{landscape}
\begin{figure}
    \centering
    \includegraphics[scale=0.5]{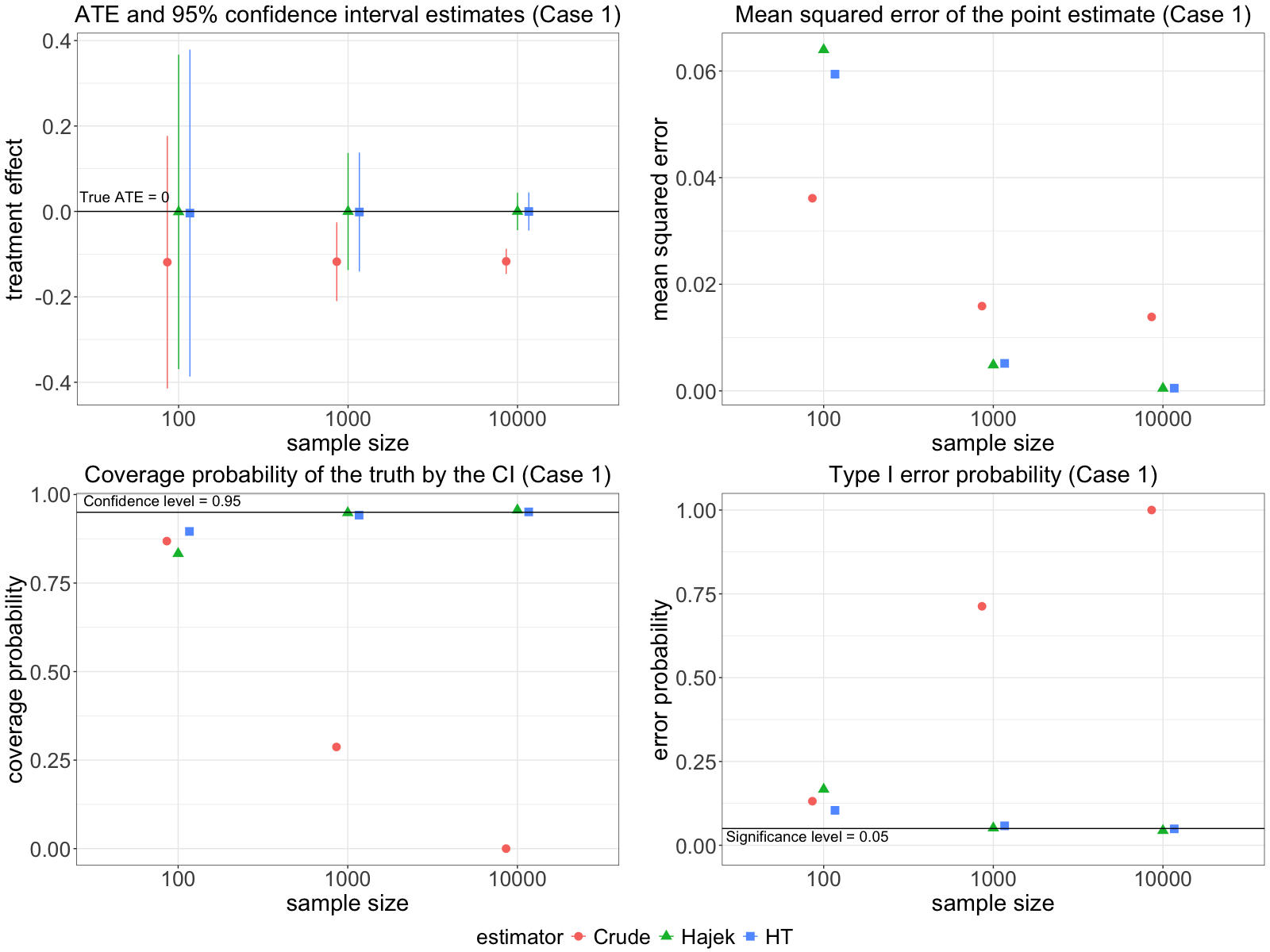}
    \caption{Case 1: simulation results of average point estimates of ATE, mean squared error, confidence intervals, coverage probability of the true ATE by the 95\% confidence interval, and the type I error probability by the crude selected-sample analysis using the selected sample and the two proposed IPW estimators under different sample sizes. The true ATE is 0 for Case 1.}\label{pic: selection_sim_case1}
\end{figure}
\end{landscape}

The crude selected-sample analysis (i.e., crude complete-case analysis) in Case 1 (red colored in Fig. \ref{pic: selection_sim_case1}) erroneously concludes a beneficial effect of the medication on reducing the risk of malaria. The $95\%$ confidence interval only contains the truth when the sample size is small due to substantial uncertainty. The coverage probability of the true ATE (a null effect of zero) by its $95\%$ confidence interval drops to zero as the sample size increases. In this null effect case, it implies that the rejection probability rises to $100\%$ as the sample size increases. The intuition is that, with a low probability of measuring the outcome among those who do not have a fever, individuals who are not assigned to the medication and do not have malaria are unlikely to be selected. Without information from these healthy individuals, the effect of the medication is overestimated, leading to selection bias. As a result, the crude selected-sample analysis yields a misleading result by exaggerating the impact of the medication. 

The results by the IPW estimators are summarized for Case 1 in Fig. \ref{pic: selection_sim_case1} (green and blue colored). Point estimates by both methods are correctly centered around zero. The mean squared error quickly drops and approaches zero. The coverage probability of the truth by the $95\%$ confidence intervals converges towards $95\%$ as the sample size increases. Equivalently, the type I error probability of rejecting the null hypothesis approaches $5\%$. Both methods are expected to fail to reject the null hypothesis at the intended size. The Hájek estimator has slightly smaller variances and narrower confidence intervals than the HT estimator.

In Case 2, consider a randomized trial of some vaccine for an infectious disease that assigns half of the individuals to the treatment group. One of the endpoints in this trial is the risk of having a headache within a certain period. The treatment ($A$) is random assignment to the vaccine or the placebo group. Its effect on the outcome of having a headache ($Y$) can be direct, through side effects that increase the risk, or indirect, mediated by the risk of infection. Whether an individual is infected ($L$) affects the chance that they report whether they had a headache ($S$) during that period. Infected individuals are more likely to have a hospital encounter and report their symptoms in terms of headache, but those uninfected are less likely to report their headache status unless contacted during follow-up. The data-generating process is 
\begin{equation*}
\begin{split}
    &A_i \sim Bernoulli(0.5)\\
    &L_i\,|\,A_i = a_i \sim Bernoulli(0.7 - 0.5a_i)\\
    &Y_i\,|\,A_i = a_i, L_i = l_i \sim Bernoulli(0.1 + 0.1a_i + 0.5l_i)\\
    &S_i\,|\,L_i = l_i \sim Bernoulli(0.1 + 0.8l_i).
\end{split}
\end{equation*}
By Corollary \ref{crlr:g_formula_Fig1}, the true ATE is $-0.15$, indicating that the vaccination on average reduces the risk of headache by $15\%$.

\begin{landscape}
\begin{figure}
    \centering
    \includegraphics[scale=0.5]{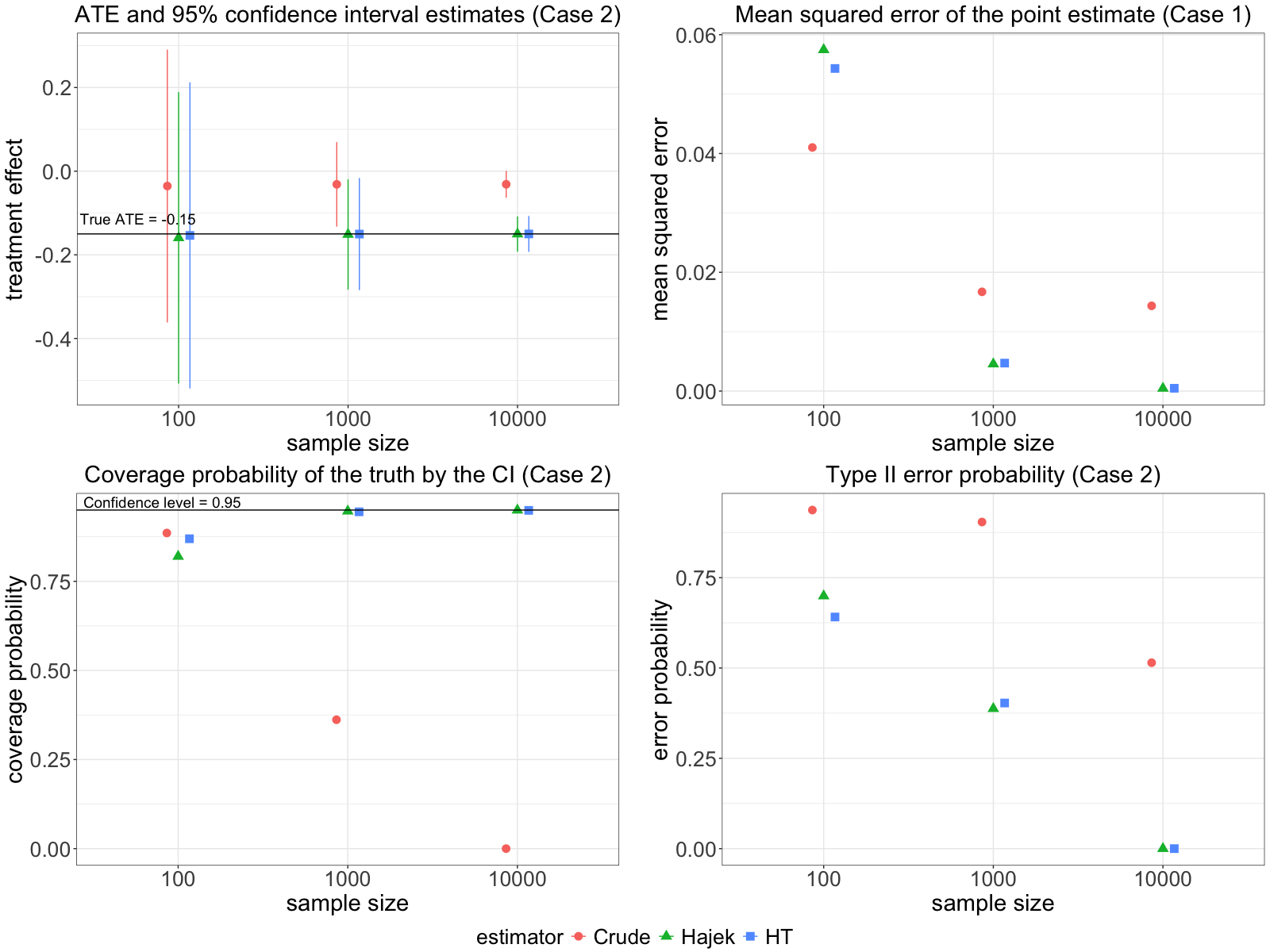}
    \caption{Case 2: simulation results of average point estimates of ATE, mean squared error, confidence intervals, coverage probability of the true ATE by the 95\% confidence interval, and the type II error probability by the crude complete-case analysis using the selected sample and the two proposed IPW estimators under different sample sizes. The true ATE is -0.15 for Case 2.}\label{pic: selection_sim_case2}
\end{figure}
\end{landscape}

The crude selected-sample analysis of Case 2 (red colored in Fig. \ref{pic: selection_sim_case2}) yields biased estimates of the ATE, with point estimates around $-0.03$. The $95\%$ confidence intervals contain zero across all three settings. Similar to Case 1, the coverage probability of the true ATE also drops to zero when the sample size increases. The rejection probability is lower than $10\%$ when the sample size is $1000$ or smaller, indicating a type II error probability higher than $90\%$. When the sample size reaches $10000$, the type II error probability is still higher than $50\%$. The explanation is that the decreased chance of reporting whether an individual has a headache among those uninfected partly masks the beneficial indirect effect of the vaccine on reducing headaches through reducing their risk of infection. Therefore, the crude selected-sample analysis fails to reveal the actual beneficial impact of the vaccination on reducing the overall risk of headaches.

Fig. \ref{pic: selection_sim_case2} shows the results of the IPW estimators for Case 2 (colored in green and blue). Both methods successfully recover the true ATE around $-0.15$ across all sample sizes. The mean squared error converges to zero. The coverage probability of the true ATE by the $95\%$ confidence interval increases to the intended level. The rejection probability increases to $100\%$ when the sample size is large enough. The power of the two methods with a sample size of $1000$ is greater than that of the crude selected-sample analysis with a sample size of $10000$ by over $10\%$.

\section{Discussion}\label{sec6}

Our work assesses and extends the utility of existing graphical rules for addressing selection bias. Grounded in the fundamental principle of reading conditional independence using d-separation, existing simple graphical rules are inherently coupled with specific identification formulas and are therefore applicable only to a subset of selection mechanisms. We show that some important existing rules \cite{bareinboim2014recovering, correa2018generalized, mathur2025simple} can impose stringent restrictions on covariate adjustment in order to avoid confounding bias while correcting for selection bias. For example, adjustment for post-treatment variables is generally discouraged under these rules, even in settings where such variables help separate selection from the outcome. While post-treatment variables should not be adjusted for when addressing confounding bias, they can nonetheless be appropriately accounted for when dealing with selection bias using methods such as IPW or the g-formula. Using two concrete examples illustrated with causal diagrams, we demonstrate the limitations of existing graphical rules and their associated identification strategies for addressing selection bias. 

For the two examples considered, we theoretically establish identifiability of the ATE using both the g-formula and IPW when relevant external information from unselected individuals is available. The g-formula accounts for the post-treatment upstream cause of selection by marginalizing over its conditional distribution, whereas IPW balances covariate distributions by modeling the probability of selection in the full population. These approaches correspond to distinct sufficient conditions that we propose for recovering the ATE. They are applicable in complex observational settings characterized by observed confounding between the treatment and post-treatment variables, as well as unobserved confounding between the upstream cause of selection and the outcome. For the IPW approach, we further propose two estimators along with corresponding variance estimators to enable statistical inference for the ATE. Through simulation studies, we evaluate their performance and show that they substantially reduce the bias introduced by conventional complete-case analyses that rely only on the selected sample.

It should be noted that our simulation studies are intentionally designed to evaluate estimator performance under settings in which the identification assumptions are satisfied, thereby confirming validity when the ATE is identifiable. We acknowledge that the assumption that observed covariates fully d-separate selection from the outcome may be strong and may not hold exactly in practice. Assessing robustness to violations of this assumption is an important direction for future research but is beyond the scope of the present work.

Beyond the two examples presented, our approach to correcting selection bias suggests new directions for recovering the ATE when conventional methods fail. First, prior research has demonstrated that the g-formula and IPW are reliable alternatives to often biased selected-sample analysis, both for addressing selection bias and for broader problems such as generalizability and transportability \cite{robins1986a, imai2010general, cole2010generalizing, stuart2011use, lesko2017generalizing, breskin2018practical, dahabreh2019generalizing, dahabreh2020extending, degtiar2023review}. Second, it is possible to adjust for post-treatment variables in ways that allow recovery of the target treatment effect without inducing overadjustment bias \cite{rosenbaum1984consequences, schisterman2009overadjustment, lu2021revisiting}. Here, “adjustment” is interpreted more broadly: the conditional distribution of a post-treatment variable may be marginalized, or the variable may be incorporated into a model for the selection process, rather than being included directly in the outcome regression.

Importantly, some previously proposed simple graphical rules focus primarily on selected-sample analysis or on estimands other than the ATE, possibly because doing so simplifies covariate selection for adjustment. In practice, however, IPW methods that leverage information beyond the selected sample have long been recognized as effective tools for addressing or mitigating selection bias and recovering the commonly targeted ATE. Selected-sample analysis (i.e., complete-case analysis) alone may be insufficient for recovering the ATE, especially in the presence of type 2 selection bias \cite{lu2022toward}. 

As noted earlier, in these cases the selection-bias and missing-data perspectives coincide: the settings are special monotone MAR problems, and our graphically derived estimators match familiar MAR estimators. A general graphical framework spanning the two perspectives, however, remains open. More broadly, the identification of causal effects in the presence of selection bias depends critically on both the target estimand of interest and the availability of external information. For instance, in our two cases the conditional average treatment effect within the selected sample is not identifiable, whereas the marginal ATE in the full population is, precisely because external information from unselected individuals is available. This highlights a key limitation of current practice: while conventional causal diagrams are useful for visualizing and diagnosing the presence of selection bias, they do not encode which variables (exposure, covariates, or outcome) are missing, nor do they specify the target estimand or whether this estimand is identifiable. Integrating causal diagrams for selection bias with graphical models for missing data (often referred to as m-graphs) \cite{mohan2021graphical, nabi2020fulllaw} offers a promising direction for overcoming these limitations.

In conclusion, selection bias remains a central challenge in epidemiologic research. Continued progress in its conceptualization, together with the development of robust identification strategies, will bring us closer to a fuller understanding of this fundamental issue.

%%%%%%%%%%%%%%%%%%%%

\textbf{Acknowledgements}

% The authors thank Forrest Crawford, Laura Forastiere, Fan Li, and Thomas Thornhill for helpful comments. 
This section has been temporarily removed from the manuscript for peer review.

\newpage
\bibliography{sn-bibliography}% common bib file

@article{hernan2004structural,
  title={A structural approach to selection bias},
  author={Hern{\'a}n, Miguel A and Hern{\'a}ndez-D{\'\i}az, Sonia and Robins, James M},
  journal={Epidemiology},
  volume={15},
  number={5},
  pages={615--625},
  year={2004},
  publisher={LWW}
}

@article{rosenbaum1983central,
  title={The central role of the propensity score in observational studies for causal effects},
  author={Rosenbaum, Paul R and Rubin, Donald B},
  journal={Biometrika},
  volume={70},
  number={1},
  pages={41-55},
  year={1983},
  publisher={Oxford University Press}
}

@article{hernan2017invited,
  title={Invited commentary: selection bias without colliders},
  author={Hern{\'a}n, Miguel A},
  journal={American journal of epidemiology},
  volume={185},
  number={11},
  pages={1048-1050},
  year={2017},
  publisher={Oxford University Press}
}

@article{bareinboim2014recovering,
  title={Recovering from selection bias in causal and statistical inference},
  author={Bareinboim, Elias and Tian, Jin and Pearl, Judea},
  journal={Proceedings of the Twenty-Eighth AAAI Conference on Artificial Intelligence},
  pages={2410-2416},
  year={2014}
}

@article{kenah2023potential,
  title={A potential outcomes approach to selection bias},
  author={Kenah, Eben},
  journal={Epidemiology},
  volume={34},
  number={6},
  pages={865-872},
  year={2023},
  publisher={LWW}
}

@article{lu2022toward,
  title={Toward a clearer definition of selection bias when estimating causal effects},
  author={Lu, Haidong and Cole, Stephen R and Howe, Chanelle J and Westreich, Daniel},
  journal={Epidemiology},
  volume={33},
  number={5},
  pages={699-706},
  year={2022},
  publisher={LWW}
}

@article{lu2024evolution,
  title={The Evolution of Selection Bias in the Recent Epidemiologic Literature—A Selective Overview},
  author={Lu, Haidong and Howe, Chanelle J and Zivich, Paul N and Gonsalves, Gregg S and Westreich, Daniel},
  journal={American Journal of Epidemiology},
  pages={kwae282},
  year={2024},
  publisher={Oxford University Press}
}

@article{schnell2024graphical,
  title={Graphical tools for detection and control of selection bias with multiple exposures and samples},
  author={Schnell, Patrick M and Kenah, Eben},
  journal={arXiv preprint arXiv:2407.20027},
  year={2024}
}

@article{correa2017causal,
  title={Causal effect identification by adjustment under confounding and selection biases},
  author={Correa, Juan and Bareinboim, Elias},
  journal={Proceedings of the AAAI Conference on Artificial Intelligence},
  volume={31},
  number={1},
  year={2017}
}

@article{correa2018generalized,
  title={Generalized adjustment under confounding and selection biases},
  author={Correa, Juan and Tian, Jin and Bareinboim, Elias},
  journal={Proceedings of the AAAI Conference on Artificial Intelligence},
  volume={32},
  number={1},
  year={2018}
}

@article{mathur2025simple,
  author  = {Mathur, Maya B. and Shpitser, Ilya},
  title   = {Simple Graphical Rules for Assessing Selection Bias in General-Population and Selected-Sample Treatment Effects},
  journal = {American Journal of Epidemiology},
  year    = {2025},
  volume  = {194},
  number  = {1},
  pages   = {267--277},
}

@article{howe2016selection,
  author  = {Howe, Chanelle J. and Cole, Stephen R. and Lau, Bryan and Napravnik, Sonia and Eron, Joseph J.},
  title   = {Selection Bias Due to Loss to Follow-Up in Cohort Studies},
  journal = {Epidemiology},
  year    = {2016},
  volume  = {27},
  number  = {1},
  pages   = {91--97},
}

@article{richardson2013single,
  title={Single world intervention graphs {SWIGs}: A unification of the counterfactual and graphical approaches to causality},
  author={Richardson, Thomas S and Robins, James M},
  journal={Center for the Statistics and the Social Sciences, University of Washington Series. Working Paper},
  volume={128},
  number={30},
  pages={2013},
  year={2013},
  publisher={Citeseer}
}

@article{pearl2014confounding,
  title={Confounding equivalence in causal inference},
  author={Pearl, Judea and Paz, Azaria},
  journal={Journal of Causal Inference},
  volume={2},
  number={1},
  pages={75-93},
  year={2014},
  publisher={De Gruyter}
}

@article{breskin2018practical,
  title={A practical example demonstrating the utility of single-world intervention graphs},
  author={Breskin, Alexander and Cole, Stephen R and Hudgens, Michael G},
  journal={Epidemiology},
  volume={29},
  number={3},
  pages={e20-e21},
  year={2018},
  publisher={LWW}
}

@article{lunceford2004stratification,
  title={Stratification and weighting via the propensity score in estimation of causal treatment effects: a comparative study},
  author={Lunceford, Jared K and Davidian, Marie},
  journal={Statistics in medicine},
  volume={23},
  number={19},
  pages={2937-2960},
  year={2004},
  publisher={Wiley Online Library}
}

@article{austin2015moving,
  title={Moving towards best practice when using inverse probability of treatment weighting (IPTW) using the propensity score to estimate causal treatment effects in observational studies},
  author={Austin, Peter C and Stuart, Elizabeth A},
  journal={Statistics in medicine},
  volume={34},
  number={28},
  pages={3661-3679},
  year={2015},
  publisher={Wiley Online Library}
}

@article{imai2010general,
  title={A general approach to causal mediation analysis},
  author={Imai, Kosuke and Keele, Luke and Tingley, Dustin},
  journal={Psychological methods},
  volume={15},
  number={4},
  pages={309},
  year={2010},
  publisher={American Psychological Association}
}

@article{hirano2003efficient,
  title={Efficient estimation of average treatment effects using the estimated propensity score},
  author={Hirano, Keisuke and Imbens, Guido W and Ridder, Geert},
  journal={Econometrica},
  volume={71},
  number={4},
  pages={1161-1189},
  year={2003},
  publisher={Wiley Online Library}
}

@book{pearl2009causality,
  title={Causality},
  author={Pearl, J},
  year={2009},
  publisher={Cambridge university press}
}

@article{reifeis2022variance,
  title={On variance of the treatment effect in the treated when estimated by inverse probability weighting},
  author={Reifeis, Sarah A and Hudgens, Michael G},
  journal={American Journal of Epidemiology},
  volume={191},
  number={6},
  pages={1092-1097},
  year={2022},
  publisher={Oxford University Press}
}

@article{cole2010generalizing,
  title={Generalizing evidence from randomized clinical trials to target populations: the {ACTG} 320 trial},
  author={Cole, Stephen R and Stuart, Elizabeth A},
  journal={American journal of epidemiology},
  volume={172},
  number={1},
  pages={107-115},
  year={2010},
  publisher={Oxford University Press}
}

@article{stuart2011use,
  title={The use of propensity scores to assess the generalizability of results from randomized trials},
  author={Stuart, Elizabeth A and Cole, Stephen R and Bradshaw, Catherine P and Leaf, Philip J},
  journal={Journal of the Royal Statistical Society Series A: Statistics in Society},
  volume={174},
  number={2},
  pages={369-386},
  year={2011},
  publisher={Oxford University Press}
}

@article{lesko2017generalizing,
  title={Generalizing study results: a potential outcomes perspective},
  author={Lesko, Catherine R and Buchanan, Ashley L and Westreich, Daniel and Edwards, Jessie K and Hudgens, Michael G and Cole, Stephen R},
  journal={Epidemiology},
  volume={28},
  number={4},
  pages={553-561},
  year={2017},
  publisher={LWW}
}

@article{dahabreh2019generalizing,
  title={Generalizing causal inferences from randomized trials: counterfactual and graphical identification},
  author={Dahabreh, Issa J and Robins, James M and Haneuse, Sebastien JP and Hern{\'a}n, Miguel A},
  journal={arXiv preprint arXiv:1906.10792},
  year={2019}
}

@article{dahabreh2020extending,
  title={Extending inferences from a randomized trial to a new target population},
  author={Dahabreh, Issa J and Robertson, Sarah E and Steingrimsson, Jon A and Stuart, Elizabeth A and Hernan, Miguel A},
  journal={Statistics in medicine},
  volume={39},
  number={14},
  pages={1999-2014},
  year={2020},
  publisher={Wiley Online Library}
}

@article{degtiar2023review,
  title={A review of generalizability and transportability},
  author={Degtiar, Irina and Rose, Sherri},
  journal={Annual Review of Statistics and Its Application},
  volume={10},
  number={1},
  pages={501-524},
  year={2023},
  publisher={Annual Reviews}
}

@article{robins1986a,
title = {A new approach to causal inference in mortality studies with a sustained exposure period—application to control of the healthy worker survivor effect},
journal = {Mathematical Modelling},
volume = {7},
number = {9},
pages = {1393-1512},
year = {1986},
author = {Robins, James M.},
}

@article{rosenbaum1984consequences,
  title={The consequences of adjustment for a concomitant variable that has been affected by the treatment},
  author={Rosenbaum, Paul R},
  journal={Journal of the Royal Statistical Society Series A: Statistics in Society},
  volume={147},
  number={5},
  pages={656-666},
  year={1984},
  publisher={Oxford University Press}
}

@article{schisterman2009overadjustment,
  title={Overadjustment bias and unnecessary adjustment in epidemiologic studies},
  author={Schisterman, Enrique F and Cole, Stephen R and Platt, Robert W},
  journal={Epidemiology},
  volume={20},
  number={4},
  pages={488-495},
  year={2009},
  publisher={LWW}
}

@article{lu2021revisiting,
  title={Revisiting overadjustment bias},
  author={Lu, Haidong and Cole, Stephen R and Platt, Robert W and Schisterman, Enrique F},
  journal={Epidemiology},
  volume={32},
  number={5},
  pages={e22-e23},
  year={2021},
  publisher={LWW}
}

@article{bezuidenhout2025swigs,
  author  = {Bezuidenhout, Dana and Forthal, Sarah and Rudolph, Kara and Lamb, Matthew R.},
  title   = {Single World Intervention Graphs (SWIGs): A Practical Guide},
  journal = {American Journal of Epidemiology},
  year    = {2025},
  volume  = {194},
  number  = {7},
  pages   = {2047--2052},
  doi     = {10.1093/aje/kwae353}
}

@inproceedings{nabi2020fulllaw,
  author    = {Nabi, Razieh and Bhattacharya, Rohit and Shpitser, Ilya},
  title     = {Full Law Identification in Graphical Models of Missing Data: Completeness Results},
  booktitle = {Proceedings of the 37th International Conference on Machine Learning (ICML 2020)},
  year      = {2020},
  volume    = {PartF16814},
  pages     = {7110--7120},
  publisher = {PMLR}
}

@article{zivich2025gcomp,
  author  = {Zivich, Paul N. and Lu, Haidong},
  title   = {Constructing g-computation estimators: two case studies in selection bias},
  journal = {Epidemiology},
  year    = {2026},
  volume  = {37},
  number  = {1},
  pages   = {50-56},
}

@article{mohan2021graphical,
  title={Graphical models for processing missing data},
  author={Mohan, Karthika and Pearl, Judea},
  journal={Journal of the American Statistical Association},
  volume={116},
  number={534},
  pages={1023--1037},
  year={2021},
  publisher={Taylor \& Francis}
}

@article{seaman2013meant,
  title={What is meant by ``missing at random"?},
  author={Seaman, Shaun and Galati, John and Jackson, Dan and Carlin, John},
  journal={Statistical Science},
  pages={257--268},
  year={2013},
  publisher={JSTOR}
}

@article{sun2018inverse,
  title={Inverse-probability-weighted estimation for monotone and nonmonotone missing data},
  author={Sun, BaoLuo and Perkins, Neil J and Cole, Stephen R and Harel, Ofer and Mitchell, Emily M and Schisterman, Enrique F and Tchetgen Tchetgen, Eric J},
  journal={American journal of epidemiology},
  volume={187},
  number={3},
  pages={585--591},
  year={2018},
  publisher={Oxford University Press}
}
\newpage

\begin{appendices}

\section{Failure of identification by existing simple graphical rules}\label{secA1}

\subsection{The selection-backdoor criterion}

The backdoor criterion was proposed to address confounding bias based on DAGs \cite{pearl2009causality}. It was extended to be the selection-backdoor criterion to assess whether conditioning on certain variables $Z$ simultaneously fixes selection bias and confounding bias \cite{bareinboim2014recovering}. The criterion conditions are as follows.

\begin{definition}[The selection-backdoor criterion \cite{pearl2014confounding}]
\label{graphical_rules1}
    Let a set $Z$ of variables be partitioned into $Z^+ \cup Z^-$ such that $Z^+$ contains all nondescendants of $A$ and $Z^-$ the descendants of $A$. $Z$ is said to satisfy the selection backdoor criterion relative to an ordered pair of variables $(A, Y)$ and an ordered pair of sets $(M, T)$ in a graph $G_s$, where $M$ are variables collected under selection bias, $P(M\,|\, S = 1)$, and $T$ are variables collected in the population-level, $P(T)$, if $Z^+$ and $Z^-$ satisfy the following conditions:
\begin{equation*}
\begin{split}
&\text{(i) $Z^+$ blocks all back door paths from $A$ to $Y$}\\
&\text{(ii) $A$ and $Z^+$ block all paths between $Z^-$ and $Y$, namely, $(Z^- \ind Y \,|\, \{A, Z^+\})$}\\
&\text{(iii) $A$ and $Z$ block all paths between $S$ and $Y$, namely, $(Y \ind S \,|\, \{A, Z\})$}\\
&\text{(iv) $Z \cup \{A, Y\} \subseteq M$, and $Z \subseteq T$}.
\end{split}
\end{equation*}
\end{definition}

The first two conditions extend the backdoor criterion to control for confounding. The third and fourth conditions allow recoverability from selection bias. If the criterion holds, 
%the effect of the treatment on the outcome is identified by the following formula:
%\begin{equation*}
%    p(y|\text{do}(a)) = \sum_z p(y|a, z, S = 1)p(z).
%\end{equation*}
the identification formula for ATE is given by
\begin{equation}
\label{formula1}
    \Delta_{\text{ATE}, 1} =  \sum_z (\E[Y\,|\,A = 1, Z = z, S = 1] - \E[Y\,|\,A = 0, Z = z, S = 1]) p(Z = z).
\end{equation}
For each of the two cases in Fig. \ref{fig:DAG_SWIG}, we consider two estimators, conditioning on $L$ ($Z = L$), or not ($Z$ empty). For both cases, when we condition on $L$, condition (ii) in the selection-backdoor criterion does not hold. This condition requires that any path between the post-treatment variable $L$, if it is conditioned on, and the outcome $Y$ must be blocked by the treatment or other pre-treatment variables. In other words, there must not be any causal paths between the post-treatment variable that we condition on, and the outcome. Alternatively, when there is no conditioning on any variable $Z$, condition (iii) in the selection-backdoor criterion, which requires any path between the outcome $Y$ and the selection $S$ must be blocked by the conditioned variables, does not hold. To summarize, by selection-backdoor criterion, conditioning on $L$ violates the strong condition for ensuring no confounding bias, while not conditioning on $L$ fails to address selection bias. The ATE is unidentified by formula (\ref{formula1}).

\subsection{The Generalized Adjustment Criterion by Correa et al.}

Following the selection-backdoor criterion based on DAGs, the Generalized Adjustment Criteria were proposed as sufficient and necessary conditions for the assertion of identification or not by certain formulas. Among them, Generalized Adjustment Criterion Type 3 (GACT3) provides the most general conditions for recovering the ATE by a flexible formula that allows the combination of external data and biased selected data \cite{correa2018generalized}.

\begin{definition}[The Generalized Adjustment Criterion Type 3 (GACT3) \cite{correa2018generalized}]
\label{graphical_rules2}
Given a causal diagram $G$ augmented with selection variable $S$, disjoint sets of variables $A$, $Y$, $Z$ and a set $Z^T \subseteq Z$; $(Z,Z^T)$ is an admissible pair relative to $A$, $Y$ in $G$ if:
\begin{equation*}
\begin{split}
&\text{(a) No element in $Z$ is a descendant in $G_{\bar{A}}$ of any $W \notin A$ lying on a proper}\\ 
&\text{causal path from $A$ to $Y$}\\
&\text{(b) All non-causal paths in $G$ from $A$ to $Y$ are blocked by $Z$ and $S$}\\
&\text{(c) $Z^T$ d-separates $Y$ from $S$ in the proper backdoor graph, where the first}\\
&\text{edge of every proper causal path from $A$ to $Y$ is removed}.
\end{split}
\end{equation*}
\end{definition}

The first two conditions control confounding bias and the last condition controls selection bias. The identification formula is
\begin{equation}
\label{formula2}
\begin{split}
        \Delta_{\text{ATE}, 2} =  \sum_z &(\E[Y\,|\,A = 1, Z = z, S = 1] - \E[Y\,|\,A = 0, Z = z, S = 1])\\
        &p(Z\backslash Z^T = z\backslash z^T\,|\,Z^T = z^T, S = 1)p(Z^T = z^T),
\end{split}
\end{equation}
where $Z$ is the full set of variables being conditioned on, and $Z^T$ is a subset of $Z$ that requires external measurement. If we condition on $Z = L$, in either of both cases, condition (a) in GACT3 is violated. If the conditioning set is empty, Case 1 violates condition (b) and (c) while Case 2 violates condition (c). It proves that the ATE can not be identified by formula (\ref{formula2}).

\subsection{Simple graphical rules under SWIG by Mathur and Shpitser}

Graphical rules based on SWIGs were also proposed for addressing selection bias \cite{mathur2025simple}. It involves two sufficient conditions to analyze whether a conditional average treatment effect (CATE) can be identified by their formula. 

\begin{definition}[The graphical rules for assessing selection bias \cite{mathur2025simple}]
\label{graphical_rules3}
Given a single-world intervention template $\mathcal{G}(a)$, the graphical rules are
\begin{equation*}
\begin{split}
&\text{(C1) } Y(a) \ind A \,|\, \{S(a), Z\}\\
&\text{(C2) } Y(a) \ind S(a) \,|\, Z.
\end{split}
\end{equation*}
\end{definition}

A premise of the two conditions in their framework is that the conditioning is on pre-treatment variables only. The first rule ensures no unadjusted confounding by conditioning on the selection and a set of adjusted covariates, and the second rule ensures no selection bias by conditioning on the same set of covariates. On its basis, a direct extension of their formula for the CATE to the ATE also follows formula (\ref{formula1}). Despite changing the graphical rules, the ATE still fails to be identified in both cases when there is no conditioning, because the second condition that tries to separate between $S(a)$ and $Y(a)$ does not hold. Identification by conditioning on $L$ is not accommodated in these simple graphical rules because $L$ is a post-treatment variable. Therefore, the ATE is not identified by formula (\ref{formula1}), again.

\section{Proof of theoretical results}\label{secA2}

\subsection{Proof of Theorem \ref{thrm:g_formula}}
\begin{proof}[Proof of Theorem \ref{thrm:g_formula}]
\begin{equation*}
\begin{split}
    & \E[Y(a)]\\ 
    = &\sum_x \E[Y(a)\,|\,X=x]p(X=x)\\
    = & \sum_x \sum_l \E[Y(a)\,|\,L(a)=l, X=x]p(L(a)=l\,|\,X=x)p(X=x)\\
    = & \sum_x \sum_l \E[Y(a)\,|\,L(a)=l, S(a)=1, X=x]p(L(a)=l\,|\,X=x)p(X=x)\\
    = & \sum_x \sum_l \E[Y(a)\,|\,L(a)=l, S(a)=1, X=x, A=a]p(L(a)=l\,|\,X=x, A=a)p(X=x)\\
    = & \sum_x \sum_l \E[Y\,|\,L=l, S=1, X=x, A=a]p(L=l\,|\,X=x, A=a)p(X=x),
\end{split}
\end{equation*}
where the third equation is by (C1) in conditions (\ref{g_formula_condition}), the fourth equation is by (C2) and (C3), and the last equation is by consistency.
\end{proof}

\subsection{Proof of Corollary \ref{crlr:g_formula_breskin}}
\begin{proof}[Proof of Corollary \ref{crlr:g_formula_breskin}]
Let $X$ be an empty set. The conditions (\ref{g_formula_condition}) hold for the SWIG in Breskin et al.\cite{breskin2018practical} where $L = W$. By Theorem \ref{thrm:g_formula}, the g-formula is reduced to
\begin{equation*}
\begin{split}
        \delta_{\text{ATE}, g} =  \sum_l &\E[Y\,|\,A = 1, L = l, S = 1]p(L=l\,|\,A=1) -\\
        &\E[Y\,|\,A = 0, L = l, S = 1]p(L=l\,|\,A=0),
\end{split}
\end{equation*}
\end{proof}

\subsection{Proof of Corollary \ref{crlr:g_formula_Fig1}}
\begin{proof}[Proof of Corollary \ref{crlr:g_formula_Fig1}]
Let $X$ be an empty set. The conditions (\ref{g_formula_condition}) hold for the SWIGs in Fig. \ref{fig:DAG_SWIG}. By Theorem \ref{thrm:g_formula}, the g-formula is reduced to
\begin{equation*}
\begin{split}
        \delta_{\text{ATE}, g} =  \sum_l &\E[Y\,|\,A = 1, L = l, S = 1]p(L=l\,|\,A=1) -\\
        &\E[Y\,|\,A = 0, L = l, S = 1]p(L=l\,|\,A=0),
\end{split}
\end{equation*}
\end{proof}

\subsection{Proof of Corollary \ref{crlr:g_formula_Fig2}}
\begin{proof}[Proof of Corollary \ref{crlr:g_formula_Fig2}]
The conditions (\ref{g_formula_condition}) hold for the SWIGs in Fig. \ref{fig:DAG_SWIG_extension} taking $X = (X_1, X_2, X_3)$ or $X = (X_1, X_2)$. By Theorem \ref{thrm:g_formula}, the g-formula is
\begin{equation*}
\begin{split}
        \delta_{\text{ATE}, g} =  \sum_x (\sum_l &\E[Y\,|\,A=1, L=l, S=1, X=x]p(L=l\,|\,A=1, X=x) -\\
        &\E[Y\,|\,A=0, L=l, S=1, X=x]p(L=l\,|\,A=0, X=x))p(X=x),
\end{split}
\end{equation*}
\end{proof}

\subsection{Proof of Theorem \ref{thrm:IPW_HT}}
\begin{proof}[Proof of Theorem \ref{thrm:IPW_HT}]
\begin{equation*}
\begin{split}
      & \quad\ \E\left[\frac{1(A_i = a, S_i = 1)Y_i}{p(S_i = 1\,|\,L_i, X_i, A_i)p(A_i = a\,|\,X_i)}\right] \\
      &= \E\left[\frac{1(A_i = a, S_i(a) = 1)Y_i(a)}{p(S_i(a) = 1\,|\,L_i(a), X_i, A_i = a)p(A_i = a\,|\,X_i)}\right]\\
      &= \E\left[\E\left[\frac{1(A_i = a, S_i(a) = 1)Y_i(a)}{p(S_i(a) = 1\,|\,L_i(a), X_i, A_i = a)p(A_i = a\,|\,X_i)}\,|\,L_i(a), X_i\right]\right]\\
      &= \E\left[\frac{\E[1(A_i = a, S_i(a) = 1)Y_i(a)\,|\,L_i(a), X_i]}{p(S_i(a) = 1\,|\,L_i(a), X_i, A_i = a)p(A_i = a\,|\,X_i)}\right]\\
      &= \E\left[\frac{\E[1(A_i = a)\,|\,L_i(a), X_i]\E[1(S_i(a) = 1)Y_i(a)\,|\,L_i(a), X_i]}{p(S_i(a) = 1\,|\,L_i(a), X_i, A_i = a)p(A_i = a\,|\,X_i)}\right]\\
      &= \E\left[\frac{\E[1(A_i = a)\,|\,L_i(a), X_i]\E[1(S_i(a) = 1)\,|\,L_i(a), X_i]\E[Y_i(a)\,|\,L_i(a), X_i]}{p(S_i(a) = 1\,|\,L_i(a), X_i, A_i = a)p(A_i = a\,|\,X_i)}\right]\\
      &= \E\left[\frac{\E[1(A_i = a)\,|\,X_i]\E[1(S_i(a) = 1)\,|\,L_i(a), X_i, A_i = a]\E[Y_i(a)\,|\,L_i(a), X_i]}{p(S_i(a) = 1\,|\,L_i(a), X_i, A_i = a)p(A_i = a\,|\,X_i)}\right]\\
      &= \E[\E[Y_i(a)\,|\,L_i(a), X_i]]\\
      &= \E[Y_i(a)].
\end{split}
\end{equation*}
The first equation is by consistency. The fourth equation is by (C2)  in conditions (\ref{IPW_HT_condition}). The fifth equation is by (C1). The sixth equation is by (C2) and (C3).
\end{proof}

\subsection{Proof of Corollary \ref{crlr:IPW_HT_breskin}}
\begin{proof}[Proof of Corollary \ref{crlr:IPW_HT_breskin}]
Let $X$ be an empty set. The conditions (\ref{IPW_HT_condition}) hold for the SWIG in Breskin et al.\cite{breskin2018practical} where $L = W$. By Theorem \ref{thrm:IPW_HT}, the IPW formula is reduced to
\begin{equation*}
\begin{split}
        \delta_{\text{ATE}, \text{IPW}} &=  \E\left[\frac{ASY}{p(S = 1\,|\,W, A)p(A=1)}\right] - \E\left[\frac{(1-A)SY}{p(S = 1\,|\,W, A)p(A=0)}\right]\\
        &= \E\left[\frac{ASY}{p(S = 1\,|\,W)p(A=1)}\right] - \E\left[\frac{(1-A)SY}{p(S = 1\,|\,W)p(A=0)}\right],
\end{split}
\end{equation*}
where the second equation is by $S \ind A | W$, i.e. there is no direct effect of A on S, for the DAG in Breskin et al.\cite{breskin2018practical}.
\end{proof}

\subsection{Proof of Corollary \ref{crlr:IPW_HT_Fig1}}
\begin{proof}[Proof of Corollary \ref{crlr:IPW_HT_Fig1}]
Let $X$ be an empty set. The conditions (\ref{IPW_HT_condition}) hold for the SWIGs in Fig. \ref{fig:DAG_SWIG}. By Theorem \ref{thrm:IPW_HT}, the IPW formula is reduced to
\begin{equation*}
\begin{split}
        \delta_{\text{ATE}, \text{IPW}} &=  \E\left[\frac{ASY}{p(S = 1\,|\,L, A)p(A=1)}\right] - \E\left[\frac{(1-A)SY}{p(S = 1\,|\,L, A)p(A=0)}\right]\\
        &= \E\left[\frac{ASY}{p(S = 1\,|\,L)p(A=1)}\right] - \E\left[\frac{(1-A)SY}{p(S = 1\,|\,L)p(A=0)}\right],
\end{split}
\end{equation*}
where the second equation is by $S \ind A | L$, i.e. there is no direct effect of A on S, for the DAG in Fig. \ref{fig:DAG_SWIG}.
\end{proof}

\subsection{Proof of Corollary \ref{crlr:IPW_HT_Fig2}}
\begin{proof}[Proof of Corollary \ref{crlr:IPW_HT_Fig2}]
The conditions (\ref{IPW_HT_condition}) hold for the SWIGs in Fig. \ref{fig:DAG_SWIG_extension}. By Theorem \ref{thrm:IPW_HT}, the IPW formula is reduced to
\begin{equation*}
\begin{split}
        \delta_{\text{ATE}, \text{IPW}} =  \E\left[\frac{ASY}{p(S = 1\,|\,L, X, A)p(A=1\,|\,X)}\right] - \E\left[\frac{(1-A)SY}{p(S = 1\,|\,L, X, A)p(A=0\,|\,X)}\right].
\end{split}
\end{equation*}
\end{proof}

\section{Steps of IPW estimation and inference}\label{secA3}

\textbf{Step 1} Two models for the treatment and selection are required. The propensity score model in the randomized trial, $p(A = 1),$ is known to us. Researchers have the flexibility to plug in the true propensity score or an estimated model from observed data. For simplicity of demonstration, we plug in the true propensity score. However, it has been shown that using the estimated propensity score model instead of the true model reduces asymptotic variances of the ATE \cite{hirano2003efficient}.

For the propensity score model for sample selection, we may fit a logistic regression model among all units, both selected ones and unselected ones, enabled by available external information on $L$, to predict the propensity of being selected by its direct upstream cause:
\begin{equation}
\label{logistic_S}
    p(S_i = 1\,|\,L_i) = \frac{e^{\beta_0 + \beta_1 L_i}}{1 + e^{\beta_0 + \beta_1 L_i}}, \quad i = 1,...,n.
\end{equation}
As we obtain estimates $\hat{\beta}_0$ and $\hat{\beta}_1$, we also get the estimated probability of being selected for each unit $\hat{p}(S_i = 1\,|\,L_i)$.

\textbf{Step 2} We compute an empirical estimate of the ATE by our choice of the IPW estimator. 
The Horvitz-Thompson (HT) IPW estimator is
\begin{equation}
\label{IPW_HT_eg}
\begin{split}
        \Delta_{\text{ATE}, \text{HT}} &= \hat{\E}[Y(1)] - \hat{\E}[Y(0)]\\
        &= \frac{1}{n}\sum_{i=1}^{n}\frac{A_iS_iY_i}{\hat{p}(S_i = 1\,|\,L_i)p(A_i=1)} - \frac{1}{n}\sum_{i=1}^{n}\frac{(1-A_i)S_iY_i}{\hat{p}(S_i = 1\,|\,L_i)p(A_i=0)}.
\end{split}
\end{equation}

An alternative IPW estimator is the Hájek estimator, which normalizes the weights to sum up to one. The formula is
\begin{equation}
\label{IPW_Hajek_eg}
\begin{split}
        \Delta_{\text{ATE}, \text{Hájek}} &= \hat{\E}[Y(1)] - \hat{\E}[Y(0)]\\
        &= \frac{\sum_{i=1}^{n}\frac{A_iS_iY_i}{\hat{p}(S_i = 1\,|\,L_i)p(A_i=1)}}{\sum_{i=1}^{n}\frac{A_iS_i}{\hat{p}(S_i = 1\,|\,L_i)p(A_i=1)}} - \frac{\sum_{i=1}^{n}\frac{(1-A_i)S_iY_i}{\hat{p}(S_i = 1\,|\,L_i)p(A_i=0)}}{\sum_{i=1}^{n}\frac{(1-A_i)S_i}{\hat{p}(S_i = 1\,|\,L_i)p(A_i=0)}}.
\end{split}
\end{equation}
In a randomized trial where $p(A_i=1)$ is a constant, the Hájek estimator does not necessarily require the input of $p(A_i=1)$ and $p(A_i=0)$ in the formula since they get canceled in computation. The Hájek estimator is also equivalent to the estimated slope from fitting a weighted least square regression of the outcome $Y$ using the treatment $A$ among all selected samples by the estimated probability of selection, $\hat{p}(S_i = 1|L_i)$. However, the variance estimate of the regression slope does not properly account for the uncertainty in estimating the weights in the first stage.

\textbf{Step 3} We calculate the variances for the two IPW estimators by the M-estimation theory \cite{lunceford2004stratification}.

\textbf{Step 3.1} The first step is to compute a system of equations \cite{reifeis2022variance}, where the parameters of interest, denoted by $\theta = (\beta_0, \beta_1, \E[Y(1)], \E[Y(0)])$, are estimated from
\begin{equation}
\label{IPW_estimating_equation}
    \sum_{i=1}^{n} \Psi(A_i, L_i, S_i, Y_i; \hat{\theta}) = \sum_{i=1}^{n} 
    \begin{pmatrix}
        \Psi_S(A_i, L_i, S_i, Y_i; \hat{\theta})\\
        \Psi_Y(A_i, L_i, S_i, Y_i; \hat{\theta})
    \end{pmatrix} = 0,
\end{equation}
where $\Psi_S(\cdot)$ denotes the subset of equations for fitting the selection model, and $\Psi_Y(\cdot)$ denotes the equations for the IPW estimator of average potential outcomes. They are derived as follows.

Suppose we fit a logistic regression (\ref{logistic_S}) for modeling selection, its estimating equations are
\begin{equation*}
    \Psi_S(A_i, L_i, S_i, Y_i; \hat{\theta}) = 
    \begin{pmatrix}
        S_i - \hat{p}(S_i = 1\,|\,L_i)\\
        L_i(S_i - \hat{p}(S_i = 1\,|\,L_i))
    \end{pmatrix}.
\end{equation*}

The estimating equations for average potential outcomes vary by the IPW estimator we choose. For the HT estimator, the equations are
\begin{equation*}
    \Psi_Y(A_i, L_i, S_i, Y_i; \hat{\theta}) = 
    \begin{pmatrix}
        \frac{A_iS_iY_i}{\hat{p}(S_i = 1\,|\,L_i)p(A_i=1)} - \hat{\E}[Y(1)]\\
        \frac{(1-A_i)S_iY_i}{\hat{p}(S_i = 1\,|\,L_i)p(A_i=0)} - \hat{\E}[Y(0)]
    \end{pmatrix}.
\end{equation*}
For the Hájek estimator, the equations are
\begin{equation*}
    \Psi_Y(A_i, L_i, S_i, Y_i; \hat{\theta}) = 
    \begin{pmatrix}
        \frac{A_iS_i(Y_i - \hat{\E}[Y(1)])}{\hat{p}(S_i = 1\,|\,L_i)p(A_i=1)}\\
        \frac{(1-A_i)S_i(Y_i - \hat{\E}[Y(0)])}{\hat{p}(S_i = 1\,|\,L_i)p(A_i=0)}
    \end{pmatrix}.
\end{equation*}

\textbf{Step 3.2} The second step is to estimate the asymptotic variance of the parameter estimates. As $n$ goes to infinity, the estimated parameters from equations (\ref{IPW_estimating_equation}) converge in distribution to a normal distribution, $\hat{\theta} \xrightarrow{d} \mathcal{N}(\theta, V(\theta)/n)$. The covariance matrix, $V(\theta)$, is estimated by a sandwich estimator
\begin{equation}
\label{sandwich_estimator}
    \begin{split}
        &\hat{V}(\hat{\theta}) = \hat{A}(\hat{\theta})^{-1} \hat{B}(\hat{\theta}) (\hat{A}(\hat{\theta})^{-1})^{T},\\
        &\hat{A}(\theta) = \frac{1}{n}\sum_{i=1}^{n}-\nabla\Psi(A_i, L_i, S_i, Y_i; \hat{\theta}),\\ 
        &\hat{B}(\hat{\theta}) = \frac{1}{n}\sum_{i=1}^{n}\Psi(A_i, L_i, S_i, Y_i; \hat{\theta})\Psi(A_i, L_i, S_i, Y_i; \hat{\theta})^T.
    \end{split}
\end{equation}

For the HT estimator, the average negative Jacobian matrix is
\begin{equation*}
    \hat{A}(\theta) = \frac{1}{n}\sum_{i=1}^{n}
    \begin{pmatrix}
        \hat{p}(S_i = 1\,|\,L_i)\hat{p}(S_i = 0\,|\,L_i) & L_i\hat{p}(S_i = 1\,|\,L_i)\hat{p}(S_i = 0\,|\,L_i) & 0 & 0\\
        L_i\hat{p}(S_i = 1\,|\,L_i)\hat{p}(S_i = 0\,|\,L_i) & L_i^2\hat{p}(S_i = 1\,|\,L_i)\hat{p}(S_i = 0\,|\,L_i) & 0 & 0\\
        \frac{A_iS_iY_i}{p(A_i=1)}\cdot\frac{\hat{p}(S_i = 0\,|\,L_i)}{\hat{p}(S_i = 1\,|\,L_i)} & \frac{A_iS_iY_iL_i}{p(A_i=1)}\cdot\frac{\hat{p}(S_i = 0\,|\,L_i)}{\hat{p}(S_i = 1\,|\,L_i)} & 1 & 0\\
        \frac{(1-A_i)S_iY_i}{p(A_i=0)}\cdot\frac{\hat{p}(S_i = 0\,|\,L_i)}{\hat{p}(S_i = 1\,|\,L_i)} & \frac{(1-A_i)S_iY_iL_i}{p(A_i=0)}\cdot\frac{\hat{p}(S_i = 0\,|\,L_i)}{\hat{p}(S_i = 1\,|\,L_i)} & 0 & 1\\
    \end{pmatrix}.
\end{equation*}
For the Hájek estimator, it is
\begin{equation*}
    \hat{A}(\theta) = \frac{1}{n}\sum_{i=1}^{n}
    \begin{pmatrix}
        \hat{p}(S_i = 1\,|\,L_i)\hat{p}(S_i = 0\,|\,L_i) & L_i\hat{p}(S_i = 1\,|\,L_i)\hat{p}(S_i = 0\,|\,L_i) & 0 & 0\\
        L_i\hat{p}(S_i = 1\,|\,L_i)\hat{p}(S_i = 0\,|\,L_i) & L_i^2\hat{p}(S_i = 1\,|\,L_i)\hat{p}(S_i = 0\,|\,L_i) & 0 & 0\\
        \frac{A_iS_i(Y_i-\hat{\E}[Y(1)])\hat{p}(S_i = 0\,|\,L_i)}{\hat{p}(S_i = 1\,|\,L_i)} & \frac{A_iS_iL_i(Y_i-\hat{\E}[Y(1)])\hat{p}(S_i = 0\,|\,L_i)}{\hat{p}(S_i = 1\,|\,L_i)} & \frac{A_iS_i}{\hat{p}(S_i = 1\,|\,L_i)} & 0\\
        \frac{(1-A_i)S_i(Y_i-\hat{\E}[Y(0)])\hat{p}(S_i = 0\,|\,L_i)}{\hat{p}(S_i = 1\,|\,L_i)} & \frac{(1-A_i)S_iL_i(Y_i-\hat{\E}[Y(0)])\hat{p}(S_i = 0\,|\,L_i)}{\hat{p}(S_i = 1\,|\,L_i)} & 0 & \frac{(1-A_i)S_i}{\hat{p}(S_i = 1\,|\,L_i)}\\
    \end{pmatrix}.
\end{equation*}

For both estimators, $\hat{B}(\hat{\theta})$ can be directly computed by plugging in their respective estimating equations (\ref{IPW_estimating_equation}).

\textbf{Step 3.3} The final step is to estimate the variance of the estimated ATE, 
\begin{equation*}
\hat{Var}(\hat{\E}[Y(1) - Y(0)]) = \hat{Var}(\hat{\E}[Y(1)]) + \hat{Var}(\hat{\E}[Y(0)]) - 2\hat{Cov}(\hat{\E}[Y(1)], \hat{\E}[Y(0)]), 
\end{equation*}
based on the asymptotic covariance matrix estimate $\hat{Cov}(\hat{\theta}) = \hat{V}(\hat{\theta})/n$.

\textbf{Step 4} Based on the asymptotic distribution, it is feasible to perform hypothesis testing and construct $100*(1-\gamma)\%$ confidence intervals by 
\begin{equation*}
    \hat{\E}[Y(1) - Y(0)] \pm Z_{1-\gamma/2}\sqrt{\hat{Var}(\hat{\E}[Y(1) - Y(0)])}.
\end{equation*}

\subsection*{Remark}

The above IPW estimators and their variance estimators can be similarly extended to observational settings in Fig. \ref{fig:DAG_SWIG_extension} by Corollary \ref{crlr:IPW_HT_Fig2}. The estimation of the propensity score model for treatment becomes a necessity. Suppose we fit a logistic regression model
\begin{equation}
\label{logistic_A}
    p(A_i = 1\,|\,X_i) = \frac{e^{\alpha_0 + \alpha_1 X_i}}{1 + e^{\alpha_0 + \alpha_1 X_i}}, \quad i = 1,...,n.
\end{equation}

For IPW estimation, we replace $p(A_i = 1)$ by $\hat{p}(A_i = 1\,|\,X_i)$ in (\ref{IPW_HT_eg}) and (\ref{IPW_Hajek_eg}) to compute the HT estimator and the Hájek estimator. The parameters to be estimated are updated by $\theta = (\alpha, \beta, \E[Y(1)], \E[Y(0)])$. Their estimating equations become
\begin{equation}
\label{IPW_estimating_equation_observational}
    \sum_{i=1}^{n} \Psi(A_i, L_i, S_i, Y_i; \hat{\theta}) = \sum_{i=1}^{n} 
    \begin{pmatrix}
        \Psi_A(A_i, L_i, S_i, Y_i; \hat{\theta})\\
        \Psi_S(A_i, L_i, S_i, Y_i; \hat{\theta})\\
        \Psi_Y(A_i, L_i, S_i, Y_i; \hat{\theta})
    \end{pmatrix} = 0.
\end{equation}
It is updated by having an extra subset of equations for the propensity score model,
\begin{equation*}
    \Psi_A(A_i, L_i, S_i, Y_i; \hat{\theta}) = 
    \begin{pmatrix}
        A_i - \hat{p}(A_i = 1\,|\,X_i)\\
        X_i(A_i - \hat{p}(A_i = 1\,|\,X_i))
    \end{pmatrix},
\end{equation*}
and replacing all $p(A_i = 1)$ by $\hat{p}(A_i = 1\,|\,X_i)$ in $\Psi_Y$. Following the definition of the sandwich estimator (\ref{sandwich_estimator}), the variance estimators can also be derived.

\end{appendices}

\end{document}